\RequirePackage{fix-cm}
\documentclass[smallextended]{svjour3}       
\smartqed  
\usepackage{graphicx}

\usepackage{mdwlist}
\usepackage[utf8]{inputenc}
\usepackage{times,natbib}
\usepackage{balance}
\usepackage{epsfig}
\usepackage{amsmath}
\usepackage{amssymb}
\usepackage{setspace}
\usepackage{multirow}
\usepackage[margin=1in]{geometry}
\usepackage{subfigure}
\RequirePackage[colorlinks,citecolor=blue,urlcolor=blue]{hyperref}
\usepackage{breakurl}
\usepackage{natbib,upgreek}
\usepackage{bbm}
\usepackage{multirow}
\usepackage{hyperref}
\usepackage{doi}
\usepackage{booktabs}
\usepackage{lineno}

\newcommand{\bftheta}{{\boldsymbol \theta}}
\newcommand{\bfs}{{\bf s}}

\begin{document}

\title{A probabilistic gridded product for daily precipitation extremes over the United States}
%


\titlerunning{A probabilistic gridded product for daily precipitation extremes}        

\author{Mark D. Risser           \and
        Christopher J. Paciorek  \and
        Michael F. Wehner        \and
        Travis A. O'Brien        \and
        William D. Collins
}


\institute{M. Risser \at
              Lawrence Berkeley National Laboratory \\
              1 Cyclotron Road, Berkeley, CA 94720
              \email{mdrisser@lbl.gov} 
}

\date{}

\maketitle

\begin{abstract}
Gridded data products, for example interpolated daily measurements of precipitation from weather stations, are commonly used as a convenient substitute for direct observations because these products provide a spatially and temporally continuous and complete source of data. However, when the goal is to characterize climatological features of extreme precipitation over a spatial domain (e.g., a map of return values) at the native spatial scales of these phenomena, then gridded products may lead to incorrect conclusions because daily precipitation is a fractal field and hence any smoothing technique will dampen local extremes. To address this issue, we  create a new ``probabilistic'' gridded product specifically designed to characterize the climatological properties of extreme precipitation by applying spatial statistical analyses to daily measurements of precipitation from the Global Historical Climatology Network over the contiguous United States. The essence of our method is to first estimate the climatology of extreme precipitation based on station data and then use a data-driven statistical approach to interpolate these estimates to a fine grid. We argue that our method yields an improved characterization of the climatology within a grid cell because the probabilistic behavior of extreme precipitation is much better behaved (i.e., smoother) than daily weather. Furthermore, the spatial smoothing innate to our approach significantly increases the signal-to-noise ratio in the estimated extreme statistics relative to an analysis without smoothing. {Finally, by deriving a data-driven approach for translating extreme statistics to a spatially complete grid, the methodology outlined in this paper resolves the issue of how to properly compare station data with output from earth system models.} We conclude the paper by comparing our probabilistic gridded product with a standard extreme value analysis of the Livneh gridded daily precipitation product. Our new data product will be made freely available on a to-be-determined public repository.

\end{abstract}

\section{Introduction} \label{section1}

Gridded data products of precipitation are a popular substitute for  daily measurements of rainfall from weather stations. The products are generated by interpolating the station measurements onto a uniform space-time grid to create a spatially and temporally complete, continuous, and homogeneous version of the raw data. For this reason, gridded products are often used to summarize the climatological properties of extreme precipitation, for example maps of return values, and then to evaluate the performance of climate models with respect to extremes. In traditional analyses of precipitation extremes using gridded products (see, e.g., \citealp{Wehner2013}, \citealp{Sylla2013}, and many others), the extreme climatology is estimated separately for each grid cell using a univariate extreme value analysis. 
However, we assert that gridded daily precipitation products are problematic data sources for constructing these extreme climatologies because daily precipitation is well-known to exhibit fractal scaling \citep[e.g.,][and numerous references therein]{Lovejoy2008,Maskey2016}, and therefore any spatial smoothing or averaging during the gridding process will diminish variability and extreme values. Additionally, a recent thread of research (\citealp{King2013}; \citealp{Gervais2014}; \citealp{Timmermans2018}) explicitly questions the appropriateness of using gridded products as a substitute for observed extremes.

As an alternative to using gridded data products, irregularly observed weather station measurements can be used to obtain spatially-complete summaries of extreme precipitation by utilizing the diverse statistics literature on spatial extreme value analysis. These statistical tools collectively analyze extremes over space for processes such as precipitation, specifically allowing one to estimate the distribution of extreme precipitation even for locations where no data is available. 
The methods broadly fall into one of four categories: max-stable processes (\citealp{deHaan1984}; \citealp{Smith1990}; \citealp{Schlather2002}; \citealp{Kabluchko2009}), which provide mathematically-founded models for  characterizing spatial dependence among extremes; copula-based approaches (\citealp{Husler1989}; \citealp{Demarta2005}; \citealp{Sang2010}; \citealp{Fuentes2013}; \citealp{Krupskii2016}), which construct a joint multivariate distribution for spatial extremes via careful modeling of transformed marginal (univariate) distributions; Bayesian methods (\citealp{Reich2012}; \citealp{Shaby2012}; \citealp{Morris2016}), which use a hierarchical framework to build numerically tractable models based upon the mathematics of max-stable processes; and nonparametric mixture models (\citealp{Gelfand2002}, \citealp{Wang2011}, \citealp{Kottas2012}), which simultaneously analyze both the average and extreme values of precipitation. See \cite{Davison2012} for an excellent review of these approaches.

In this paper, we seek to characterize the climatology of extreme precipitation based on measurements from approximately five thousand Global Historical Climatology Network (GHCN) stations (see Section~\ref{section2}) over the contiguous United States (CONUS). Unfortunately, practically speaking, many of the aforementioned statistical methods are only suitable for a small number of weather stations. For example,  \cite{Davison2012} {use} a data set with just 36 stations, \cite{Saunders2017} employ 173 stations to derive a max-stable process to model extreme precipitation in Australia, and \cite{Shaby2012} analyze the largest data set considered to date (to our knowledge)  comprised of approximately one thousand stations. Regardless of the number of weather stations of interest, a second limitation of the aforementioned methods is that they are only appropriate for homogeneous spatial domains. CONUS, on the other hand, is a highly heterogeneous spatial domain with variable topography interacting with a diverse set of physical phenomena that produce extreme precipitation. These phenomena include atmospheric rivers on the west coast, tropical cyclones on the east coast, and mesoscale convective systems in the Great Plains.

{In light of these shortcomings in existing methodologies for spatial extremes, note that we do not need to simulate realistic fields of seasonal maxima, which is a complicating factor in the aforementioned max-stable and Bayesian approaches to spatial extreme value analysis (furthermore, note that we do not intend to model the full space-time distribution of daily precipitation, as in \citealp{serinaldi2014simulating}). Instead, we simply wish to characterize the statistics of extreme precipitation over CONUS. In this case,} 
one apparently viable option would be to employ latent variable or conditional independence models (\citealp{Cooley2007}; \citealp{CraigmileGuttorp2013}; \citealp{Mannshardt2013}), which provide flexible methodologies suitable for larger data sets and heterogeneous spatial domains. These methods assume that daily precipitation totals over space occur independently of one another and are conditional on latent processes that characterize spatial dependence in extremes. The assumption of independence invalidates the application of latent variable methods to modeling  extremes of an atmospheric process such as precipitation because daily precipitation measurements at nearby stations are certainly not independent. This dependence is induced by the spatial coherence of storm systems so that if one weather station experiences an extreme on a specific day it is more likely that a nearby station will also experience an extreme on the same day. Failing to account for this so-called ``storm dependence'' means that the resulting maps of extreme statistics will be misleading since the true spatial signals will be commingled with noise.

In order to create a viable alternative analytical technique, we  first conduct an extreme value analysis to estimate the extreme statistics of precipitation for each station and then interpolate these statistics to a fine grid. In order to account for spatial coherence in these extreme statistics, we use second-order nonstationary spatial Gaussian process models that account for heterogeneities  in the climatology of extreme precipitation and enable statistical inference at locations where we do not have observations of daily precipitation. By construction, this framework accounts for both first-order heterogeneities in the expected values due to topography and second-order heterogeneities in the co-variability. 
Then, we use a nonparametric, or ``block,'' bootstrap approach to characterize uncertainty in the extreme statistics of precipitation while accounting for storm dependence.  Intuitively, our approach can be seen as commuting the order of operations relative to existing analytical methods that in general first apply spatial and temporal interpolation to station data onto a regular grid and then calculate extreme statistics from that gridded product. We demonstrate that our method yields a larger signal-to-noise ratio in estimates of extreme statistics relative to the ratios yielded by more traditional analytical methods that do not borrow strength spatially.

Focusing on daily measurements of precipitation from the Global Historical Climatology Network (GHCN) over the contiguous United States (CONUS), the result of our analysis is a probabilistic gridded product that describes the spatial climatological distribution of extreme precipitation. To illustrate why our methodology provides an improved characterization of extreme precipitation statistics, consider two nearby towns in a homogeneous region, e.g., the Great Plains. While these two towns might experience different storms for any given day or month, our general hypothesis is  that these two towns will exhibit similar statistics of extreme precipitation because of their similar climatology. The same argument holds over the rest of CONUS, especially after correcting for the effects of elevation. As a result, it makes more sense to borrow strength over space and conduct smoothing on the statistics of extremes rather than daily weather itself. A traditional extreme value analysis using gridded daily products does not incorporate this important borrowing of strength over space.

{Two threads of research should be mentioned in relation to the novel methodology described in this paper. The first is \cite{diaconescu2015remapping} and references therein, which explores the order of operations when regridding data products and climate model output for comparison. In this paper, they compare a ``first-step'' procedure (which first interpolates daily weather fields and then computes a derived index on the common grid) and a ``last-step'' procedure (which first computes the derived index and then interpolates to a common grid),  finding that the last-step procedure yields much smaller errors from the regridding process (a result that is robust to the interpolation method). The intuition for our approach supports a preference for the last-step procedure and is indeed quite similar, except that (1) we propose re-gridding station data instead of preprocessed data products, and (2) we interpolate parameters from a statistical model instead of data summaries (e.g., the largest seasonal daily precipitation total): hence, our results immediately yield a variety of gridded climatological summaries (e.g., return levels or return probabilities for any return period) instead of requiring a separate interpolation scheme for each desired summary.} {Second, our methodology is related to regional frequency analysis (RFA) (\citealp{wallis2007regional}; see also \citealp{norbiato2007regional} and \citealp{yang2010regional}) which, as with the work at hand, seeks to provide spatially-complete maps of the climatology of extreme precipitation based on the interpolation of pointwise summary statistics calculated for each of a set of weather stations. However, the methodologies are quite different: the cornerstone of RFA is that ``data from sites within a homogeneous region can be pooled to improve the reliability of the magnitude-frequency estimates'' (\citealp{wallis2007regional}), and hence RFA requires the specification of such regions. This is a nontrivial task, particularly for a large, heterogeneous domain such as CONUS. Our approach, on the other hand, allows the data itself to determine the degree of local homogeneity in the climatology of extreme precipitation across a set of weather stations, and is hence more easily applied to a generic spatial domain.}

{Finally, we note that the methodology outlined in this paper resolves an outstanding problem for evaluation of {Earth System Models (ESMs)} with respect to extremes: namely, how to properly compare irregularly observed station data with output from {ESMs}. {Comparison of ESMs against station data poses several major challenges. First, the volume of the atmosphere sampled by station data is typically orders of magnitude smaller than the volumes represented by a typical ESM grid cell.  While ESM sub-grid parameterizations are designed to emulate the parent distributions of measurable quantities and to report averages across these distributions, stations can record samples from any part of the parent distribution of their measurable quantities, including the ``tails.''  Second, the typical integrations of ESMs forced just with external boundary conditions are designed to mimic the climatological statistics of typical weather conditions for each surface station, but not the precise deterministic time-series of the weather actually recorded by those stations.  Thus, the question is how best to compare deterministic time series of point measurements against a statistical characterization of the mesoscale climate conditions consistent with those measurements.  The methodology developed in this paper addresses this question by framing the model-data intercomparison in terms of the extreme statistics from the outset, and by developing a method for interpolating statistical properties between stations that produces mesoscale and synoptic-scale statistics directly comparable to those from an ESM.}}

The paper proceeds as follows: in Section~\ref{section2} we introduce the GHCN data used to create our probabilistic gridded product, and in Section~\ref{section3} we describe the extreme value analysis, spatial smoothing, and block bootstrap approach. In Section~\ref{section4}, we present the results of our analysis and give a summary of the probabilistic gridded product, and in Section~\ref{section5} we conduct a comparative analysis using the Livneh gridded daily precipitation product. Section~\ref{section6} concludes the paper.

\section{Data} \label{section2}

The data used for the following analysis consist of daily measurements of total precipitation (in millimeters) obtained from the Global Historical Climatology Network (GHCN; \citealp{ghcnd_data} and 
\citealp{Menne2012}) over the contiguous United States (CONUS). 
The GHCN is quite extensive over CONUS, consisting of over twenty thousand weather stations with measurements dating back to the late 19th century, although of course the length and quality of individual records are highly variable. In addition to daily precipitation measurements, the database includes three quality control flags, providing for each day an indication of the data quality (``QFLAG''), source quality (``SFLAG''), and measurement quality (``MFLAG''). Quality control for the raw (nonmissing) daily data values involved the following criteria: (1) values were kept only if the QFLAG field was blank, meaning the measurement did not fail any quality assurance checks; (2) values were kept only if the SFLAG field was not equal to ``S'' (which implies that the measurement may differ significantly from ``true'' daily data); (3) any values with MFLAG equal to ``T'' were set to a measurement of 0 mm (``T'' indicates a ``trace of precipitation''). After processing the daily values based on the quality control flags, we selected the subset of stations that had a minimum of 66.7\% nonmissing daily precipitation measurements over December 1, 1949 through November~30, 2017. This procedure resulted in a high-quality set of daily precipitation measurements for $n = 5202$ stations (see Figure~\ref{GHCNstations}) covering 68 years.

To establish some notation, define $Z_{tk}(\bfs_i)$ to be the precipitation measurement, in mm, for day $k = 1, \dots, m_t$ in a fixed three-month season (e.g., December, January, and February or DJF) of year $t = 1950, \dots, 2017$ at station $\bfs_i \in \mathcal{S} = \{\bfs_1, \dots, \bfs_n\} \subset G$, where $m_t$ is the number of daily observations in a fixed season of year $t$, $\mathcal{S}$ denotes the $n=5202$ stations shown in Figure~\ref{GHCNstations}, and $G$ denotes the contiguous United States. Note that the year represents a ``season year'' such that, for example, the 1950 DJF is defined as December, 1949 to February, 1950. 

\section{Statistical methods} \label{section3}

Recall from Section~\ref{section1} that the essence of our method is to first obtain estimates of the climatological features of extreme precipitation based on station data (Section~\ref{section31}) and then use a spatial statistical approach to infer the true underlying climatology over a fine grid (Section~\ref{section32}), accounting for uncertainty and storm dependence using a nonparametric block bootstrap approach (Section~\ref{section33}). We now provide specific details on each step of this analysis.

\subsection{Stage 1: extreme value analysis for individual stations} \label{section31}

The following provides a framework for modeling extreme value statistics of daily precipitation in a fixed three-month season (i.e., DJF, MAM, JJA, or SON) over CONUS, first considering each station individually. While there are several different ways to characterize the extreme values of a stochastic process (see, e.g., \citealp{Coles2001}), we opt for the generalized extreme value (GEV) family of distributions, which is a modeling framework for the maxima of a process over a pre-specified ``block,'' here, each three-month season. For an arbitrary station $\bfs$, define the seasonal maximum in year $t$ as $Y_t(\bfs)$, that is, $Y_t(\bfs) = \max_k \{ Z_{tk}(\bfs) : k = 1, \dots, m_t \}$. \cite{Coles2001} (Theorem~3.1.1, page~48) shows that (for large $m_t$) the cumulative distribution function (CDF) of $Y_t(\bfs)$ is a member of the GEV family
\begin{equation} \label{gev_fam}
G_{\bfs, t}(y) \equiv \mathbb{P}(Y_t(\bfs) \leq y) = \exp\left\{-\left[ 1 + \xi_t(\bfs)\left(\frac{y - \mu_t(\bfs)}{\sigma_t(\bfs)}\right) \right]^{-1/\xi_t(\bfs)} \right\}, 
\end{equation}
defined for $\{ y: 1 + \xi_t(\bfs)(y - \mu_t(\bfs))/\sigma_t(\bfs) > 0 \}$. The GEV family of distributions~(\ref{gev_fam}) is characterized by three space-time specific parameters: the location parameter $\mu_t(\bfs) \in \mathcal{R}$ (which describes the center of the distribution), the scale parameter $\sigma_t(\bfs)>0$ (which describes the spread of the distribution), and the shape parameter $\xi_t(\bfs) \in \mathcal{R}$. The shape parameter $\xi_t(\bfs)$ is the most important in terms of determining the qualitative behavior of the distribution of daily rainfall at a given location: if $\xi_t(\bfs)<0$, the distribution has a finite upper bound; if $\xi_t(\bfs) >0$, the distribution has no upper limit; if $\xi_t(\bfs) = 0$, the distribution is again unbounded and the CDF~(\ref{gev_fam}) is interpreted as the limit $\xi_t(\bfs) \rightarrow 0$ (\citealp{Coles2001}). 

The GEV framework is commonly referred to as the ``block maxima'' approach to extreme value analysis. The point process or ``peaks over threshold'' (POT) approach is often  preferred to the block maxima approach because, as in the threshold excess model, estimates of the climatological coefficients are obtained from all extreme values (i.e., those that exceed a high threshold) instead of the single maximum over a block of time. But, as discussed in \cite{Coles2001} section~4.3.1, the main challenge is identifying a threshold for what is considered ``extreme": too small a threshold violates the mathematical (asymptotic) basis of the POT approach, leading to biased coefficient estimates, while too large a threshold results in a very small number of exceedances, yielding large variance. Also, in practice, when conducting station-specific extreme value analyses, the POT approach resulted in a larger number of numerical optimization errors. Finally, in this case where we wish to characterize trends in extremes over time, it is not clear that a temporally-constant threshold is appropriate. Therefore, we opted to use the GEV framework for this analysis. However, as a sensitivity analysis we also conducted the full analysis using the POT approach and found that results were similar (see Figure~\ref{figureC4} for a comparison).

As the notation in~(\ref{gev_fam}) suggests, we can specify flexible time-varying models for the spatially-varying climatological coefficients. For the analysis in this paper, we use a simple temporal trend, where the location parameter varies linearly with time and the other coefficients are constant in time:
\begin{equation} \label{coef_model}
\mu_t(\bfs) = \mu_0(\bfs) + \mu_1(\bfs) t, \hskip2ex \sigma_t(\bfs) \equiv \sigma(\bfs), \hskip2ex \xi_t(\bfs) \equiv \xi(\bfs)
\end{equation}
(following, e.g., \citealp{Westra2013} and others). We henceforth refer to $\mu_0(\bfs)$, $\mu_1(\bfs)$, $\sigma(\bfs)$, and $\xi(\bfs)$ as the \textit{climatological coefficients} for location $\bfs$, as these values describe the climatological distribution of extreme precipitation in each year. Note that the trend model~(\ref{coef_model}) averages over both inter-annual variability (e.g., the El Ni\~no/Southern Oscillation or ENSO) and lower frequency modes of variability (e.g., the Pacific Decadal Oscillation or PDO), such that we only attempt to characterize temporally smooth trends in extremes. Some authors (e.g., \citealp{Cooley2007}; \citealp{Risser2017}) have also permitted the scale parameter to contain covariates allowing the width of the GEV distribution to co-vary. Uncertainty in the magnitude of the shape parameter is generally quite large, negating any benefits of allowing it to co-vary. 
In this paper, however, we consider only~(\ref{coef_model}) because in a statistical sense it performed as well  as more sophisticated trend models for individual stations. 

\sloppypar{
Considering all years, the statistical model (or log-likelihood) for all of the observed seasonal maxima at station $\bfs$, defined as ${\bf y}(\bfs) = \{ y_{t}(\bfs) : t = 1950, \dots, 2017 \}$, conditional on the climatological coefficients $\mu_0(\bfs)$, $\mu_1(\bfs)$, $\sigma(\bfs)$, and $\xi(\bfs)$, is
\begin{equation} \label{likelihood}
\mathcal{L}\big(\mu_0(\bfs), \mu_1(\bfs), \sigma(\bfs), \xi(\bfs); {\bf y}(\bfs) \big) = -\sum_{t = 1950}^{2017}\log \sigma(\bfs) \hskip35ex
\end{equation}
\[
\hskip24ex - \big[1 + 1/\xi(\bfs)\big] \sum_{t = 1950}^{2017} \log \left[ 1+ \xi(\bfs)\left(\frac{y_t(\bfs) - [\mu_0(\bfs) + \mu_1(\bfs)t]}{\sigma(\bfs)} \right)\right]
\]
\[
\hskip20ex - \sum_{t = 1950}^{2017}  \left[ 1+ \xi(\bfs)\left(\frac{y_t(\bfs) - [\mu_0(\bfs) + \mu_1(\bfs)t]}{\sigma(\bfs)} \right)\right]^{-1/\xi(\bfs)}.
\]
The fact that~(\ref{likelihood}) involves a sum indicates that we assume independence across years, which is a reasonable assumption given the time-varying statistical model in~(\ref{coef_model}).
}

While the spatially-varying climatological coefficients in~(\ref{coef_model}) are of interest themselves, we are often more interested in a summary of the climatological coefficients known as the $r$-year return value (sometimes called the return level). The $r$-year return value, denoted $\phi^{(r)}_t(\bfs)$, is defined as the seasonal maximum daily precipitation total (i.e., $Y_t(\bfs) = \max_k \{ Z_{tk}(\bfs) \}$) that is expected to be exceeded on average once every $r$ years. In other words, $\phi^{(r)}_t(\bfs)$ is an estimate of the $1-\frac{1}{r}$ quantile of the distribution of seasonal maximum daily precipitation in year $t$ at station $\bfs$, i.e.,
$P\big(Y_{t}(\bfs) > \phi^{(r)}_t(\bfs)\big) = {1}/{r}$.
Because of how the year-specific distribution~(\ref{gev_fam}) has been defined, $\phi^{(r)}_t(\bfs)$ can be written in closed form in terms of the climatological coefficients:
\begin{equation} \label{returnVal}
\phi^{(r)}_t(\bfs) = \left\{ \begin{array}{ll}
[\mu_0(\bfs) + \mu_1(\bfs)t] - \frac{\sigma(\bfs)}{\xi(\bfs)}\big[1 - \{-\log(1-1/r)\}^{-\xi(\bfs)}\big],  & \xi(\bfs) \neq 0 \\[1ex]
[\mu_0(\bfs) + \mu_1(\bfs)t] - \sigma(\bfs) \log\{-\log(1-1/r)\},  & \xi(\bfs) = 0
\end{array} \right. 
\end{equation}
(\citealp{Coles2001}). Furthermore, while the return value is the extreme quantile of the extreme value distribution in each year, we can equivalently calculate the {return period} for a particular magnitude event $x$ in year $t$, denoted $\rho_t^{(x)}(\bfs)$, which indicates that there is a one in $\rho_t^{(x)}(\bfs)$ chance that an event at least as large as $x$ will occur in year $t$ at location $\bfs$. In other words, $\rho_t^{(x)}(\bfs)$ is the inverse probability of the seasonal maximum daily precipitation total in year $t$ exceeding $x$, i.e.,\linebreak
$P\big(Y_{t}(\bfs) > x\big) = {1}/{\rho_t^{(x)}(\bfs)}$.
The return period is available in closed form by inverting~(\ref{returnVal}):
\begin{equation} \label{returnPer}
\rho_t^{(x)}(\bfs) = \left\{ \begin{array}{ll}
\left(1 - \exp\left\{- \big[1- \xi(\bfs)([\mu_0(\bfs) + \mu_1(\bfs)t] - x)/\sigma(\bfs) \big]^{-1/\xi(\bfs)} \right\} \right)^{-1},  & \xi(\bfs) \neq 0 \\[1ex]
\left(1 - \exp\left\{- \exp\big\{ ([\mu_0(\bfs) + \mu_1(\bfs)t] - x)/\sigma(\bfs) \big\} \right\} \right)^{-1},  & \xi(\bfs) = 0
\end{array} \right. 
\end{equation}
(\citealp{Coles2001}). 

In summary, this first stage of our method involves applying the GEV analysis defined by~(\ref{coef_model}) and~(\ref{likelihood}) to the observed seasonal maxima, independently for each station. For this step, we use the {\tt climextRemes} software package (\citealp{R_climextRemes}), which is an R/Python package for conducting extreme value analysis with climate data. We use {\tt climextRemes} functionality to obtain maximum likelihood estimates of the climatological coefficients, denoted $\{ \widehat{\mu}_0(\bfs), \widehat{\mu}_1(\bfs), \widehat{\sigma}(\bfs), \widehat{\xi}(\bfs) \}$ for $\bfs \in \mathcal{S}$, which are estimates of the true climatological coefficients $\{ {\mu}_0(\bfs), {\mu}_1(\bfs), {\sigma}(\bfs), {\xi}(\bfs) \}$.

\subsection{Stage 2: spatial statistical modeling for the climatological coefficients} \label{section32}

In order to account for dependence across the climatological coefficients over the spatial domain $G$, we use second-order nonstationary spatial Gaussian process models for each of the spatially-varying climatological coefficients in~(\ref{coef_model}). Gaussian processes (GPs) are an extremely popular tool in statistical modeling, and are broadly applied in spatial and environmental statistics as well as machine learning and emulation of complex mathematical models. GPs are a form of ``nonparametric'' or nonlinear regression, as they characterize a nonlinear relationship between a set of inputs and an output: in our application, the inputs are geographic coordinates and the output (or response) is one of the estimated climatological coefficients $\{ \mu_0, \mu_1, \log\sigma, \xi \}$. Intuitively, GPs interpolate or smooth the output variable to infer a nonlinear relationship with the inputs.

The GP is a popular choice for modeling point-referenced, continuously-indexed spatial processes like the climatological coefficients because all finite-dimensional distributions are known to be Gaussian (see Equation~\ref{theta_dist} below), and because the GP is completely specified by a characterization of its first- and second-order properties. Furthermore, the second-order properties can be specified using any valid spatial covariance function to describe the degree and nature of spatial dependence (or smoothness) in the environmental process of interest. For example, define $\{ u(\bfs): \bfs \in G \}$ to be a general process defined over a spatial domain $G \subset \mathbb{R}^2$; without loss of generality, suppose $u(\cdot)$ is mean-zero. The spatial covariance function of $u(\cdot)$ is defined as
\[
C(\bfs, \bfs') \equiv \text{Cov}\big[u(\bfs), u(\bfs')] = E[u(\bfs)u(\bfs')]
\]
for all $\bfs, \bfs' \in G$, where $E[\cdot]$ is the expected value with respect to the distribution of $u(\cdot)$. The covariance function is always symmetric (i.e., $C(\bfs, \bfs') = C(\bfs', \bfs)$) and must be a nonnegative definite function. In practice, $C$ is often assumed to be second-order stationary, meaning that the covariance is fully defined by the separation vector $\bfs - \bfs'$. 
However, for the application at hand, this is a highly restrictive assumption, because the spatial covariance for environmental processes likely varies across the domain. Therefore, we use a \textit{nonstationary} covariance function, which allows the covariance to depend on location and can account for second-order heterogeneities over the spatial domain of interest. A GP that uses a nonstationary covariance function is often referred to simply as a nonstationary GP. For further details on this subject, we refer the interested reader to \cite{Risser2016}.

A nonstationary Gaussian process can be used in a statistical model for the climatological coefficients over CONUS as follows. Let $\theta$ represent an arbitrary coefficient, i.e., $\theta \in \{ \mu_0, \mu_1, \log\sigma, \xi \}$ (we model the scale parameter $\sigma$ on the logarithmic scale because $\sigma$ must be positive); the Gaussian process model for each $\theta(\bfs)$ can be framed as the statistical linear model
\begin{equation} \label{theta_mod}
\theta(\bfs) = \delta_\theta(\bfs) + u_\theta(\bfs), \hskip3ex \bfs \in G.
\end{equation}
In~(\ref{theta_mod}), $\delta_\theta(\bfs) \equiv E[\theta(\bfs)]$ represents the expected or average behavior of $\theta(\bfs)$ and $u_\theta(\bfs)$ represents a mean-zero residual process that characterizes deviations between $\theta(\bfs)$ and $\delta_\theta(\bfs)$. In a more traditional linear regression setting (i.e., ordinary least squares), $u_\theta(\bfs)$ is assumed to be (spatially) independent, but here we instead model $u_\theta(\bfs)$ as a spatially correlated nonstationary Gaussian process with a mean fixed at zero and covariance function $C_\theta$. For a fixed set of locations $\mathcal{S} = \{\bfs_1, \dots, \bfs_n\} \subset G$, (\ref{theta_mod}) implies that
\begin{equation} \label{theta_dist}
\bftheta \equiv \big(\theta(\bfs_1), \dots, \theta(\bfs_n) \big) \sim N_n \big( \boldsymbol{\delta}_\theta, {\bf \Sigma}_\theta \big),
\end{equation}
where $N_d({\bf a}, {\bf B})$ denotes a $d$-variate Gaussian distribution with mean vector ${\bf a}$ and covariance matrix ${\bf B}$. In~(\ref{theta_dist}), $\boldsymbol{\delta}_\theta = \big(\delta_\theta(\bfs_1), \dots, \delta_\theta(\bfs_n) \big)$ and the $i,j$ element of ${\bf \Sigma}_\theta$ is $C_\theta(\bfs_i, \bfs_j)$. 

To complete the model specification in~(\ref{theta_mod}), we must specify the form of $\delta_\theta(\bfs)$ as well as the covariance function $C_\theta$; for full details, see Appendix \ref{appendixA}. In short, the mean function $\delta_\theta(\bfs)$ will be linear in an elevation-based covariate with a spatially-varying effect, and the covariance function for $C_\theta$ is second-order nonstationary such that $C_\theta(\bfs, \bfs') \neq C_\theta(\bfs - \bfs')$. Again, intuitively, using a nonstationary covariance function allows the second order properties of $u_\theta(\cdot)$ to vary over space, including both variance and the magnitude and  direction of the spatial extent of the smoothing. A statistical treatment of~(\ref{theta_dist}) uses a likelihood-based approach (e.g., maximum likelihood estimation) to estimate the statistical parameters associated with the mean and covariance.

\subsection{Nonparametric bootstrap for uncertainty quantification} \label{section33}

Finally, we specify a systematic framework for combining the GEV likelihood in~(\ref{likelihood}) and the Gaussian process priors for the coefficient fields. However, note that we have not yet specified how to interrelate the likelihood~(\ref{likelihood}) across stations, and these relationships are a critical component of a statistical model for daily precipitation extremes over space. As mentioned in Section~\ref{section1}, the simplest approach is the conditional independence or latent variable approach (see, e.g., \citealp{Cooley2007}, \citealp{CraigmileGuttorp2013}, and \citealp{Mannshardt2013}), which assumes independence across stations in the likelihood and relies on the Gaussian process priors to capture dependence in extremes. However, while this approach is somewhat feasible for large data sets and heterogeneous spatial domains, the fact that it does not account for storm dependence makes it theoretically incorrect. And, as mentioned in Section~\ref{section1}, existing approaches for more appropriately modeling extremes over space are insufficient for daily precipitation measurements from a large network like GHCN over CONUS. 

Alternatively, we propose a hierarchical framework that combines the practical benefits of the conditional independence approach with a more appropriate characterization of the uncertainty in the resulting estimates of the climatological coefficients. First, recall from Section~\ref{section31} that we have maximum likelihood estimates of the coefficients, denoted $\widehat{\bftheta} =  \big( \widehat{\theta}(\bfs_1), \dots, \widehat{\theta}(\bfs_n) \big)$ (where again $\theta \in \{ \mu_0, \mu_1, \log\sigma, \xi \}$). This vector is an estimate of the true underlying spatial field ${\bftheta} =  \big( {\theta}(\bfs_1), \dots, {\theta}(\bfs_n) \big)$. However, because these estimates are obtained independently for each station, $\widehat{\bftheta}$ includes  true spatial signal, spatial noise from storm dependence, and any additional measurement noise. The following hierarchical model links the estimates $\widehat{\bftheta}$ with $\bftheta$:
\begin{equation}\label{hierTheta}
\begin{array}{c}
\widehat{\bftheta} \big| \bftheta \sim N_n( \bftheta, {\bf D}_\bftheta)\\
\bftheta \sim N_n \big( \boldsymbol{\delta}_\theta, {\bf \Sigma}_\theta \big)
\end{array}
\end{equation}
following, e.g., \cite{Holland2000}, \cite{Tye2015}, and \cite{Russell2016}. This model specifies that the estimates of $\widehat{\bftheta}$ conditional on the true field $\bftheta$ are unbiased with covariance ${\bf D}_\bftheta$, and that the true signal $\bftheta$ follows a Gaussian process as defined in Section~\ref{section32}. In~(\ref{hierTheta}), ${\bf D}_\bftheta$ quantifies the discrepancy between $\widehat{\bftheta}$ and $\bftheta$, which we model as a diagonal matrix. In other words, we assume that the error in $\widehat{\bftheta}$ is spatially-independent. The spatial covariance ${\bf \Sigma}_\theta$, on the other hand, is a non-diagonal matrix that characterizes spatial coherence in $\bftheta$. Estimates of the process mean $\widehat{\boldsymbol{\delta}}_\theta$, spatial covariance $\widehat{\bf \Sigma}_\theta$, and spatially-independent error $\widehat{\bf D}_\bftheta$ are obtained via local likelihood techniques (\citealp{RisserCalder2017}; for more details see Appendix \ref{appendixA}) using the marginalized model
\begin{equation} \label{margMod}
\widehat{\bftheta} \sim N_n(\boldsymbol{\delta}_\theta, {\bf \Sigma}_\theta + {\bf D}_\bftheta),
\end{equation} 
where we have integrated out the process ${\bftheta}$ from the joint distribution implied by~(\ref{hierTheta}). 

The Gaussian process assumption allows us to recover an estimate of the true process $\theta(\cdot)$ at the station locations, denoted $\widetilde{\bftheta} = \big(\widetilde{\theta}(\bfs_1), \dots, \widetilde{\theta}(\bfs_n) \big)$. Conditional on $\widehat{\bftheta}$ as well as maximum likelihood estimates $\widehat{\boldsymbol{\delta}}_\theta$, $\widehat{\bf \Sigma}_\theta$, and $\widehat{\bf D}_\bftheta$, our best estimate of the true climatological coefficient process is
\begin{equation} \label{predMean0}
\widetilde{\bftheta} = \widehat{\boldsymbol{\delta}}_\theta + \widehat{\bf \Sigma}_\theta \big[ \widehat{\bf \Sigma}_\theta  + \widehat{\bf D}_\bftheta\big]^{-1}\big( \widehat{\bftheta} - \widehat{\boldsymbol{\delta}}_\theta \big),
\end{equation}
which is also known as the \textit{kriging predictor} in a traditional geostatistical framework. Here, $\widehat{\bf \Sigma}_\theta \big[ \widehat{\bf \Sigma}_\theta  + \widehat{\bf D}_\bftheta\big]^{-1}$ is the matrix version of [signal]/[signal + noise], so we can see that best estimate $\widetilde{\bftheta}$ is the sum of the spatial mean ($\widehat{\boldsymbol{\delta}}_\theta$) and a spatial residual term ($ \widehat{\bftheta} - \widehat{\boldsymbol{\delta}}_\theta$) that is re-scaled based on the relative magnitude of the signal and noise. In fact, this relationship can be generalized to a generic location $\bfs' \in G$ for which we do not have observations of daily precipitation (e.g., on a fine grid). Similar to (\ref{predMean0}), our best estimate of the true climatological coefficient at $\bfs'$ is
\begin{equation} \label{predMean}
\widetilde{\theta}(\bfs') = \widehat{\delta}_\theta(\bfs') + \widehat{\bf c}^\top_{\bftheta, \theta(\bfs')} \big[ \widehat{\bf \Sigma}_\theta  + \widehat{\bf D}_\bftheta\big]^{-1}\big( \widehat{\bftheta} - \widehat{\boldsymbol{\delta}}_\theta \big).
\end{equation}
In~(\ref{predMean}), $\widehat{\delta}_\theta(\bfs')$ is the estimated mean at $\bfs'$ and 
$\widehat{\bf c}^\top_{\bftheta, \theta(\bfs')} =  \big( \widehat{C}_\theta(\bfs_1, \bfs'), \dots, \widehat{C}_\theta(\bfs_n, \bfs') \big)$
is the estimated covariance between $\theta(\cdot)$ at $\bfs'$ and values at the station locations. After obtaining best estimates for each of the climatological coefficients, we can then calculate best estimates of the return values $\widetilde{\phi}^{(r)}_{t}(\bfs')$ using~(\ref{returnVal}) and the return periods $\widetilde{\rho}^{(x)}_{t}(\bfs')$ using~(\ref{returnPer}). 

However, assuming that ${\bf D}_\bftheta$ is spatially-independent (i.e., diagonal) implies that the spatial covariance ${\bf \Sigma}_\theta$ describes both the true spatial signal as well as any spatially-correlated error from storm dependence. Therefore, our best estimates of the climatological coefficients $\widetilde{\theta}(\bfs)$ from~(\ref{predMean}) contain both real spatial signal and spatially-correlated noise from the storm dependence. To account for this issue, we use the nonparametric or ``block" bootstrap to characterize uncertainty in our estimates of the climatological coefficients as well as return values/periods. The block bootstrap requires no assumptions of independence within each year of data and preserves the spatial and temporal features of the data by re-sampling entire years of data, using the same resampled years for all station locations. We rely on the bootstrapping procedure to address storm dependence, since any real spatial signal will show up in most of the bootstrap data sets. 
The block bootstrap approach proceeds as follows: define ${\bf y}(\bfs_i) = \{ y_{t}(\bfs_i) : t = 1950, \dots, 2017 \}$ to be the observed vector of the seasonal maxima for station $i$. Then, for $b=1, \dots, B$, the bootstrap sample is obtained by drawing $T = 68$ years from $\{ 1950, \dots, 2017 \}$, denoted $\{a_1^*, a^*_2, \dots,a^*_{T} \}$, so that the $b$th bootstrap sample for station $i$ is
\[
{\bf y}^*_{bi} = \big( y_{a_1^*}(\bfs_i), y_{a_2^*}(\bfs_i), \dots, y_{a_T^*}(\bfs_i)  \big)
\]
For each bootstrap sample, we use the multi-stage procedure outlined in Sections \ref{section31} and \ref{section32} with (\ref{predMean}) to obtain the best estimates of the coefficients $\widetilde{\theta}_b(\bfs)$, return values $\widetilde{\phi}^{(r)}_{b,t}(\bfs)$, and return periods $\widetilde{\rho}_{b,t}^{(x)}(\bfs)$. The resulting field of bootstrap standard errors for any quantity of interest, e.g., the $r$-return value estimate in year $t$, is calculated as 
\[
v^{(r)}_j(\bfs) = \sqrt{ \frac{1}{B-1} \sum_{b = 1}^B \left(\widetilde{\phi}^{(r)}_{b,t}(\bfs) - \frac{1}{B}\sum_{b = 1}^B \widetilde{\phi}^{(r)}_{b,t}(\bfs) \right)^2}.
\]

A potentially cleaner way to handle the storm dependence would be to include spatially-correlated error in ${\bf D}_\theta$, allowing us to explicitly separate the true spatial signal from the error. Exploratory analyses (see Appendix \ref{appendix0}) reveal the presence of spatially-correlated error in the coefficient estimates $\widehat{\bftheta}$ at non-negligible scales due to storm dependence. As such, we considered using a correlated (non-diagonal) empirical estimate for the covariance ${\bf D}_\bftheta$ in~(\ref{hierTheta}) (for example, \citealp{Holland2000} use a bootstrap-based estimate). However, this approach resulted in several problems. First, it is extremely difficult to estimate a high-dimensional covariance matrix (here, a $5202 \times 5202$ matrix) based on bootstrap sampling from a limited temporal record. Second, in this case we found that the signal and noise have similar spatial scales (again see Appendix \ref{appendix0}), making it difficult to appropriately separate signal from noise. Third, we were not confident that including correlated error resulted in the correct amount of shrinkage, because the resulting uncertainty estimates from a test case were unrealistic. 


\section{Results} \label{section4}

\subsection{Spatial statistics and uncertainty quantification} \label{ssec:spatialanduq}

\begin{figure}[!t]
\begin{center}
\includegraphics[trim={0 0 0 0mm}, clip, width = \textwidth]{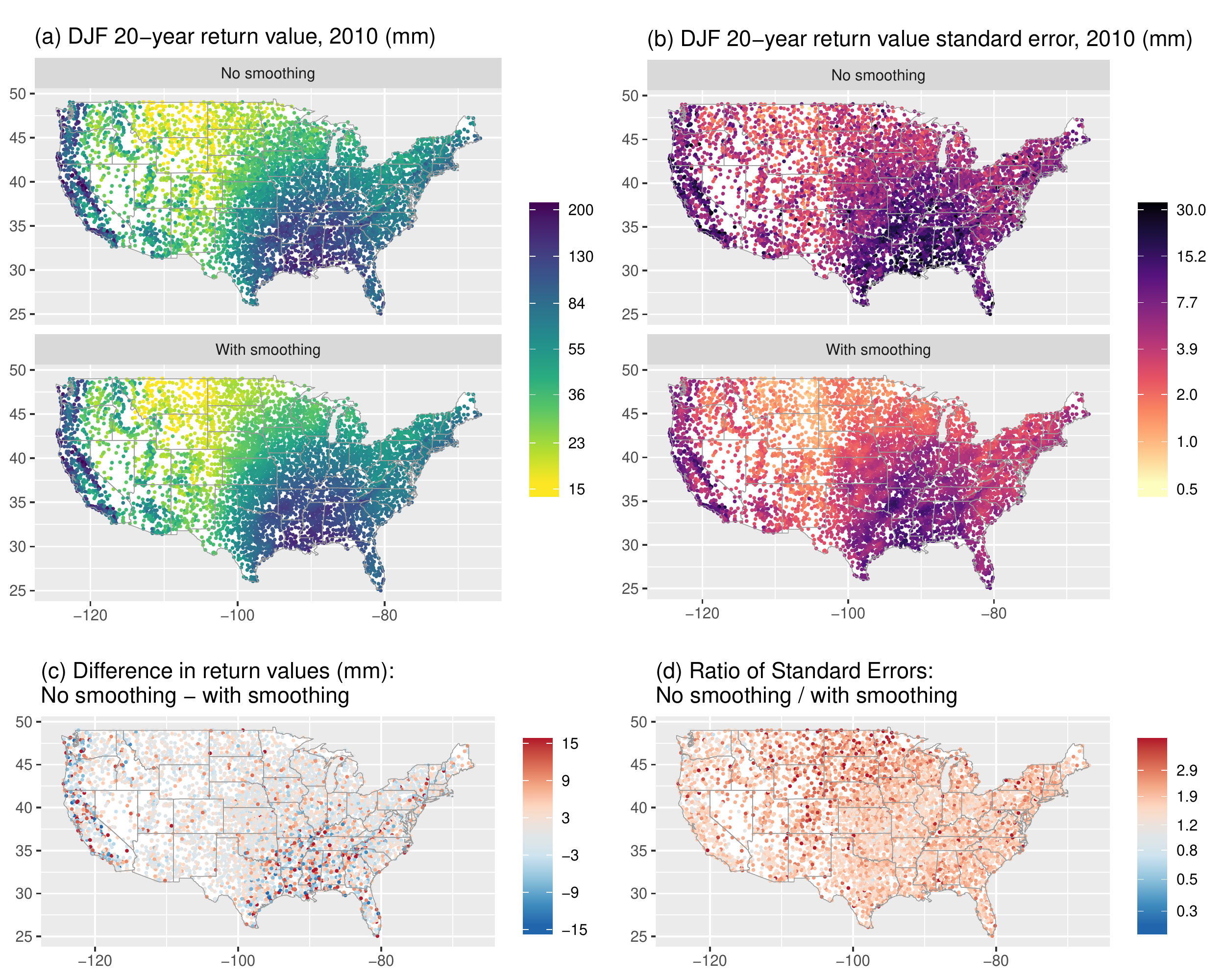}
\caption{Comparison of 20-year return values for DJF in 2017 (mm; panel a) and bootstrap standard errors (mm; panel b) for a traditional analysis with no spatial smoothing versus our approach with spatial smoothing. An explicit comparison of the return values and standard errors are shown in panel (c) and (d), respectively.}
\label{figure1}
\end{center}
\end{figure}

While the original goal of this analysis was to create an improved probabilistic gridded product, the methodology yields an additional benefit of giving increased confidence in the extreme statistics of precipitation at the stations. To illustrate this concept, we first present maps of the estimated 20-year return value for DJF in 2010 at the GHCN station locations (Figure~\ref{figure1}; estimated return values for the other seasons are shown in Figure~\ref{figureC2} in the Appendix). The year 2010 was somewhat arbitrarily chosen, although importantly 2010 can be directly compared with existing data products (e.g., Livneh; see Section~\ref{section5}). For comparison, we present results from a ``traditional'' extreme value analysis, where the return values are estimated independently for each station, i.e., without spatial smoothing, alongside estimates obtained from our new method. Figure~\ref{figure1} also shows the bootstrap standard errors in the estimated return values, again with and without spatial smoothing, as well as a direct comparison of the difference in return values and ratio of standard errors. The two approaches yield similar return values (i.e., the ``signal''), with differences of less than several millimeters (except in areas with large return values, e.g., the southeast US). However, when considering the bootstrap standard errors (an estimate of the ``noise''), our new approach with smoothing yields much smaller standard errors across much of the domain. In other words, spatial smoothing yields approximately the same signal but reduced noise. The bottom right panel of Figure~\ref{figure1} explicitly highlights this reduction in noise by plotting the ratio of standard errors with and without smoothing: for DJF, smoothing results in uncertainty estimates that are on average approximately half as large, with as much as a three-fold reduction in the upper Great Plains. The results are similar for the other seasons (see Figure~\ref{figureC3}), with the largest reduction in uncertainty occurring in JJA.

\begin{figure}[!t]
\begin{center}
\includegraphics[trim={0 0 0 0mm}, clip, width = \textwidth]{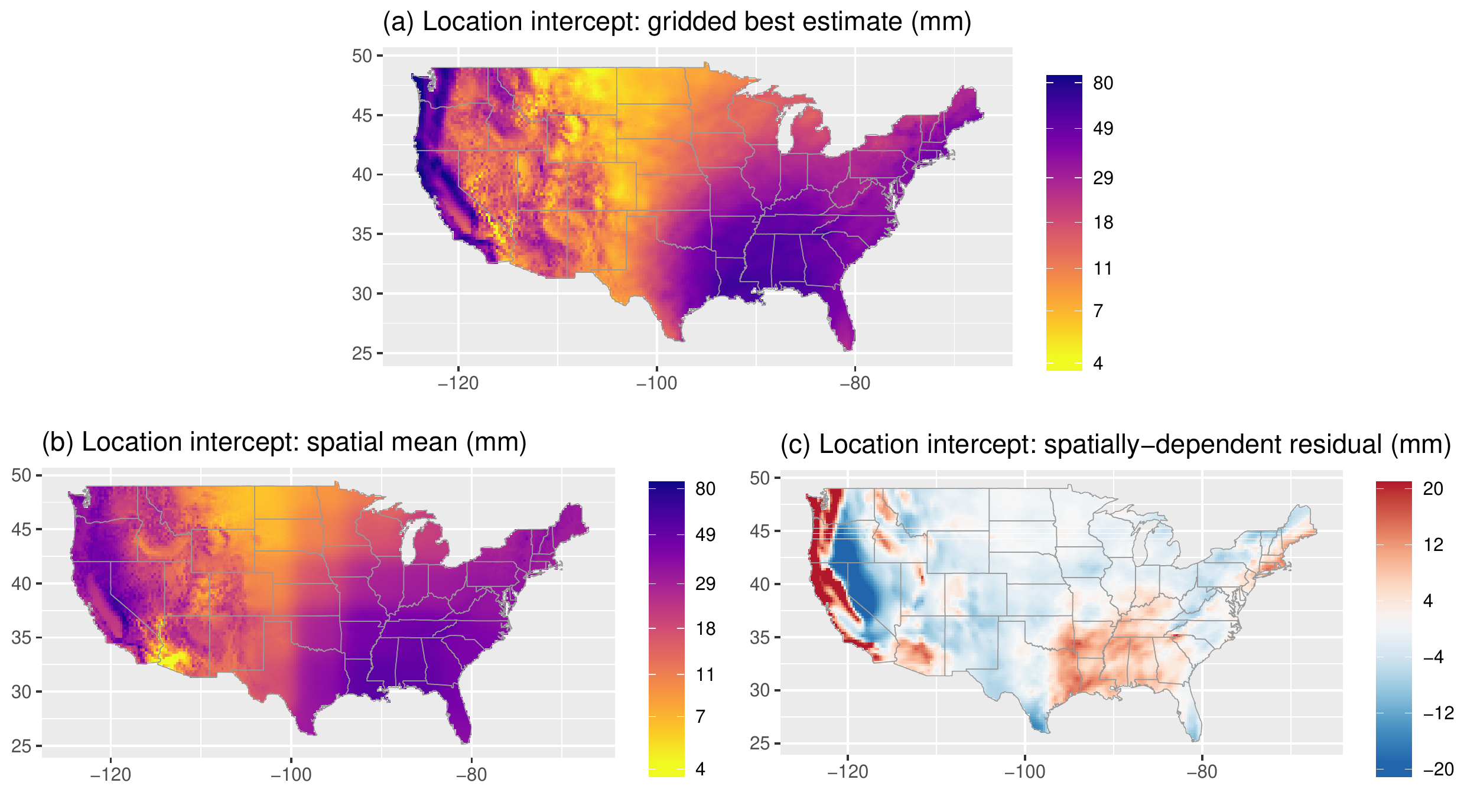}
\caption{Gridded best estimates of the location intercept $\mu_0(\bfs)$ over CONUS (mm) for DJF (panel~a), decomposed into the elevation-based spatial mean (panel~b) and the spatially-dependent residual (panel~c). (In other words, panel~(a) is the sum of panels~(b) and~(c).)}
\label{figure2}
\end{center}
\end{figure}

\begin{figure}[!t]
\begin{center}
\includegraphics[trim={0 25 0 0mm}, clip, width = 0.9\textwidth]{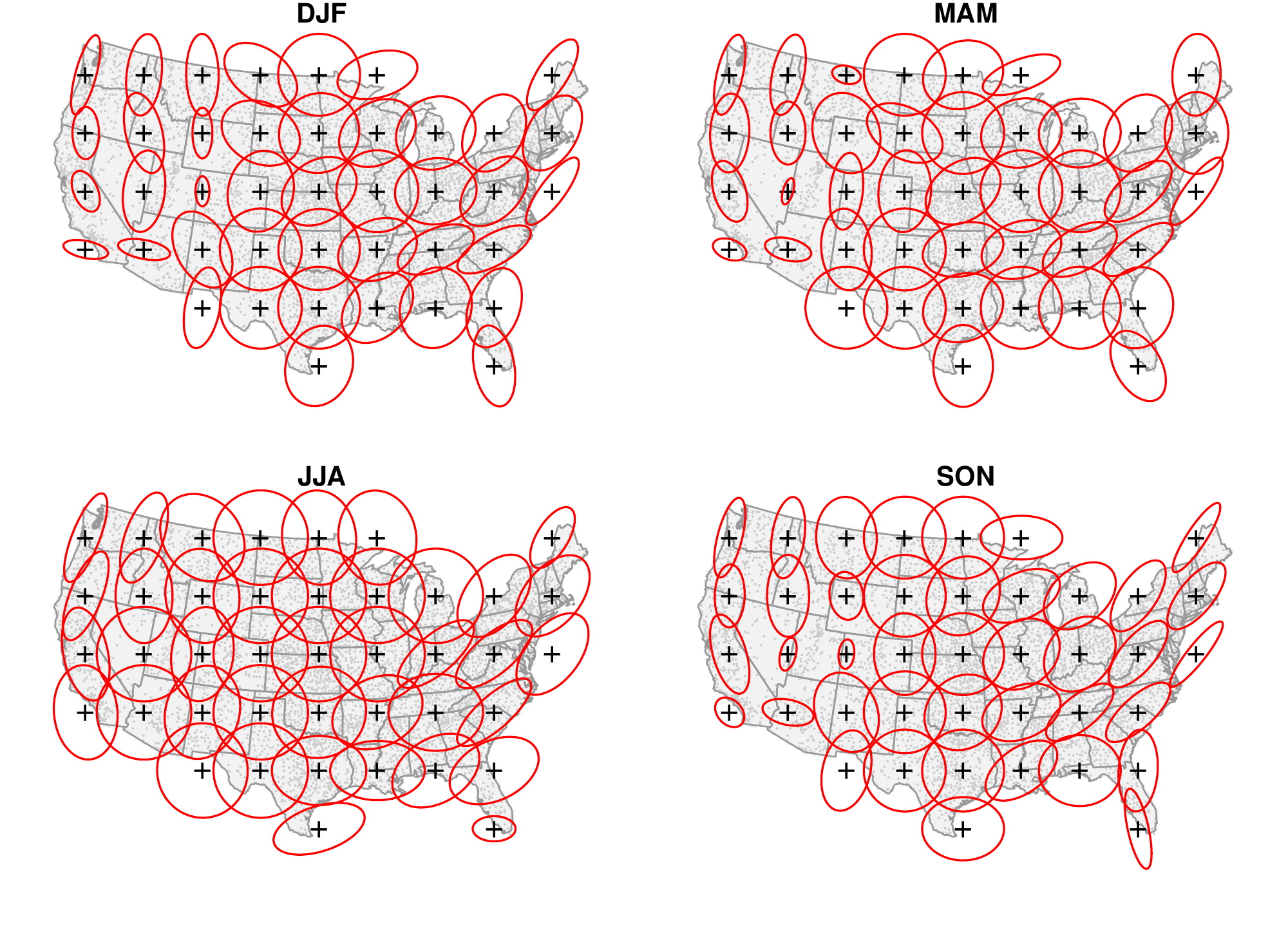}
\caption{Directional spatial length-scale for the location intercept $\mu_0(\bfs)$ for each season, estimated empirically from the data. The ellipses are a heuristic representation of the magnitude and direction of spatial dependence in $\mu_0(\bfs)$.}
\label{figure3}
\end{center}
\end{figure}

In addition to the increased signal-to-noise ratio, the use of spatial statistics in our novel approach yields insight into the behavior of extreme precipitation. As described in Section~\ref{section32}, nonstationary Gaussian process models allow the spatial length-scale (i.e., magnitude and direction of spatial dependence) to vary across the spatial domain. As an illustration, we show results for the location intercept coefficient $\mu_0(\bfs)$ in DJF, which characterizes the center of the extreme value distribution (constant over time) for each spatial location and hence drives the magnitude of the return values. Our best estimates of $\mu_0(\bfs)$ in DJF are shown in Figure~\ref{figure2}(a), where we show statistically-gridded estimates (using Equation~\ref{predMean}) as it is easier to visualize the spatial distribution with a ``filled-in'' map. Figure~\ref{figure2} also shows the elevation-based spatial mean ($\widehat{\delta}_{\theta}(\bfs')$ in Equation~\ref{predMean}) in Figure~\ref{figure2}(b), as well as the spatially-dependent residual ($\widehat{\bf c}^\top_{\bftheta, \theta(\bfs')} \big[ \widehat{\bf \Sigma}_\theta  + \widehat{\bf D}_\bftheta\big]^{-1}\big( \widehat{\bftheta} - \widehat{\boldsymbol{\delta}}_\theta \big)$ in Equation \ref{predMean}) in Figure~\ref{figure2}(c). (Corresponding plots for MAM, JJA, and SON are shown in the Appendix Figures \ref{figure2b}, \ref{figure2c}, and \ref{figure2d}.) The spatial mean in Figure~\ref{figure2}(b) represents the estimated first-order properties of $\mu_0(\bfs)$, and characterizes a linear relationship between elevation and the center of the extreme value distribution. The use of a spatial Gaussian process yields the spatially-dependent residual term, which characterizes additional nonlinear relationships over space in $\mu_0(\bfs)$ that are not captured by the elevation covariate. 

Focusing on the spatially-dependent residual, first note that the residuals are smooth over the southeast United States and upper Great Plains but highly heterogeneous across the Rocky Mountains and along the west coast. This is not surprising, but note that these differences can be explicitly characterized by the nonstationary Gaussian process, which estimates this length-scale directly from the data -- and, these differences persist even after accounting for elevation directly via the spatial mean in Figure~\ref{figure2}(b). (Note that elevation artifacts can still be seen in Figure~\ref{figure2}(c), indicating some inadequacy in our characterization of the elevation-based mean.) To visualize these differences, consider Figure~\ref{figure3}, which shows a heuristic representation of the spatially-varying length scale (both magnitude and direction) via a set of ellipses, which are estimated locally across CONUS at the mixture component locations. Intuitively, a long/skinny ellipse (e.g., coastal Washington state in DJF) means that the length scale of spatial dependence is much larger in one direction than in the orthogonal direction; a circular ellipse (e.g., western Illinois in DJF) means that the length-scale is roughly the same in all directions. Similarly, a large ellipse corresponds to long-range spatial dependence while a small ellipse corresponds to shorter-range spatial dependence. Note that the Gaussian process estimates translate to the variations in heterogeneity noted in the gridded spatial residuals: in DJF, the Rocky mountain range and west coast generally has a shorter length-scale (i.e., highly heterogeneous) while the southeast US and upper Great Plains generally have longer length-scales (i.e., highly smooth or homogeneous). There are other interesting features to these ellipse maps, e.g., coastal effects (both east and west), as well as the southwest-to-northeast orientation of the ellipses across the southeast US (particularly in SON). Also, note the seasonal differences in the spatial length-scale, particularly DJF vs. JJA, which could be due to either the climatology or the different storm types leading to extreme precipitation in winter and summer. 

To be clear, estimating a spatially-varying length scale directly from the data reiterates the importance of using a statistical approach to gridding the GEV coefficients and hence the return values: the data itself informs the degree of smoothing, which varies over the domain. This approach is more physically meaningful than heuristic approaches like bilinear interpolation or inverse-distance weighting, or even ``ordinary kriging'' using an isotropic Gaussian process, which uses a constant (and circular) spatial length-scale.

\begin{figure}[!t]
\begin{center}
\includegraphics[trim={0 0 0 0mm}, clip, width = \textwidth]{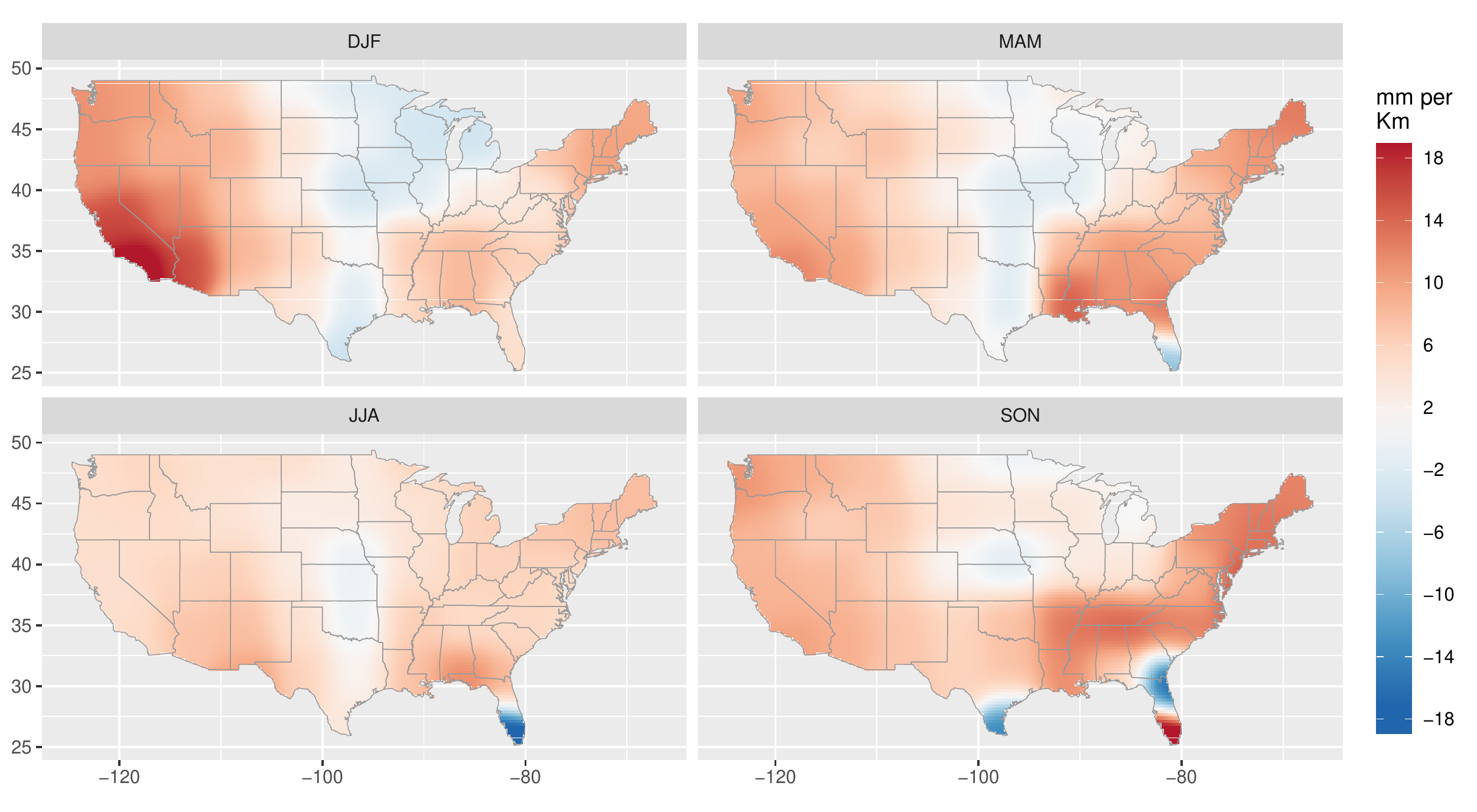}
\caption{Spatially-varying elevation correction for the Gaussian process mean of $\mu_0(\bfs)$, across each season (mm of precipitation per Km of elevation). Positive values indicate larger return values for higher elevations; negative values indicate smaller return values for higher elevations.}
\label{figure4}
\end{center}
\end{figure}

As mentioned in Section~\ref{section32}, we include an orographic correction in the nonstationary Gaussian process mean for each GEV coefficient. Intuitively, this orographic correction involves estimating a linear relationship between elevation and each GEV coefficient empirically from the data, while allowing the magnitude and sign of this relationship to vary smoothly across CONUS. While the mean is linear in elevation, note that this does not completely specify the relationship between extreme precipitation and elevation because of the spatial smoothing induced by the Gaussian process (this is illustrated in Figure~\ref{figure2}). Figure~\ref{figure4} shows the data-driven, spatially-smoothed estimate of the relationship between extreme precipitation and elevation, across each season. Dark red areas indicate a strong positive relationship relationship between elevation and extreme precipitation (i.e., higher elevation corresponds to stronger storms), while blue areas indicate a negative relationship. Across all of the seasons, the relationship between elevation and extreme precipitation is generally positive, with the largest effects showing up in the western United States. Of course, elevation itself is rather homogeneous over much of the central United States. However, in the western United States (where elevation plays an important role in the climatology of precipitation), there are important seasonal differences in the relative elevation/precipitation relationship: for example, especially in California, the effect is larger for DJF than JJA. This is not surprising, as there is a strong wet/dry seasonal cycle in the western United States, but it is encouraging that the nonstationary GP was able to infer this directly from the data.

Of course, many gridded daily products also include an orographic correction, most notably the Parameter-elevation Relationships on Independent Slopes Model (PRISM; see \citealp{Daly2008} and the references therein). While our orographic correction is applied to the climatological coefficients rather than daily precipitation itself, several notes of comparison should be made. First of all, like PRISM, our approach involves a linear relationship between the climate variable and elevation (compare Equation 2 in \citealp{Daly2008} with Equation \ref{eq:spatMean} in the Appendix). However, PRISM proceeds to incorporate a distance weighting scheme (see Equation 3 in \citealp{Daly2008}) that is fixed \textit{a priori} and spatially constant (albeit elevation-dependent). Our approach with nonstationary Gaussian processes, on the other hand, estimates the appropriate length scale for distance weighting directly from the data, using a non-constant weighting scheme over space. Furthermore, the Gaussian process prediction (via Equation~\ref{predMean}) implicitly accounts for the over-representation issue in distance-weighted averaging (such as in PRISM) while also explicitly removing noise in the data due measurement error. Finally, it is worth restating that the most important difference between our approach and PRISM is in \emph{what} is actually being smoothed; whereas PRISM smooths precipitation itself, our method smooths GEV distribution parameters.

\subsection{Probabilistic gridded product}

\begin{figure}[!t]
\begin{center}
\includegraphics[trim={0 0 0 0mm}, clip, width = \textwidth]{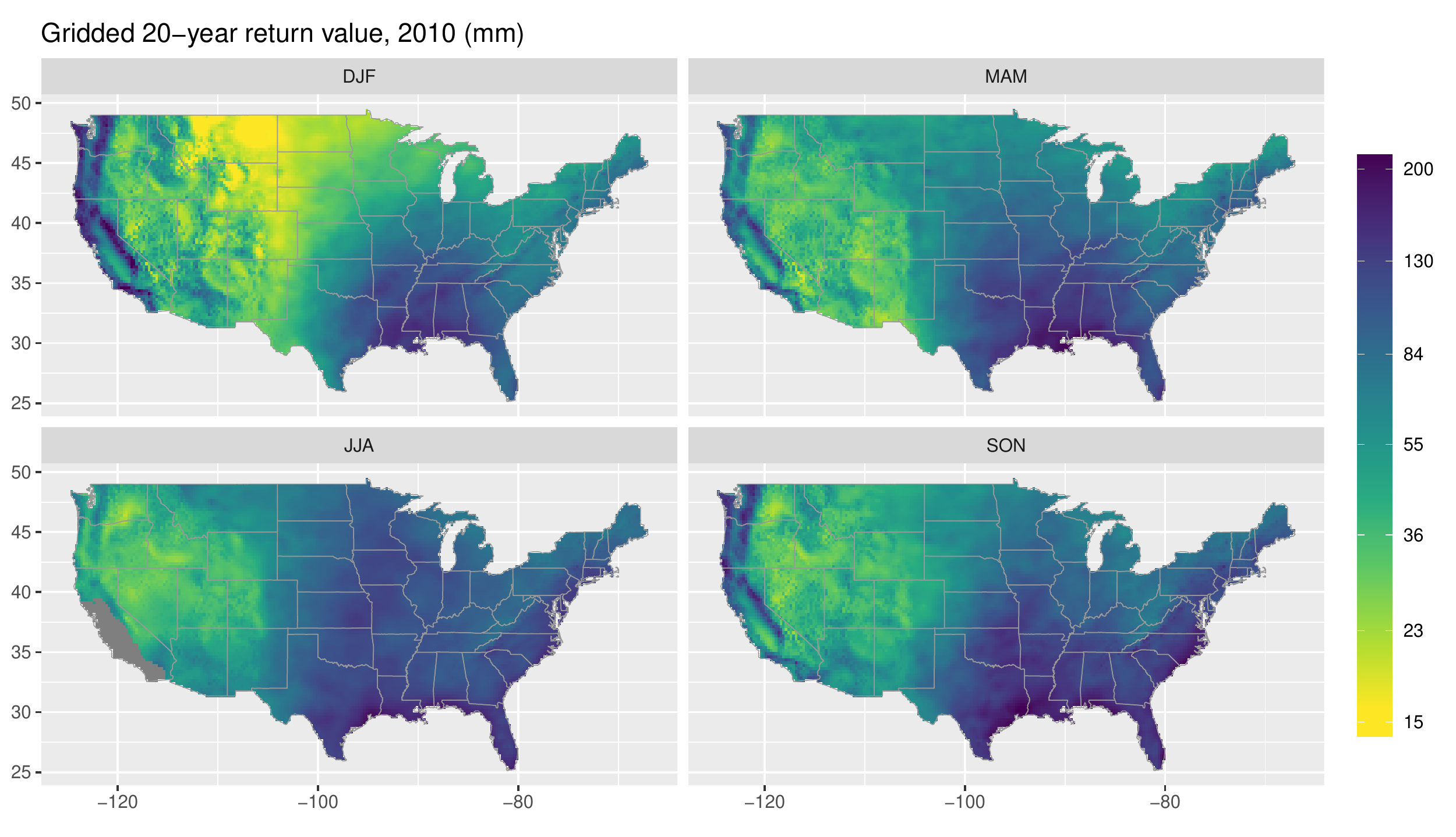}
\caption{Spatially-complete 20-year return values (mm) for CONUS, across each season. Note: part of California is masked out for JJA because the dry season in this region means that the stations in this area do not register any ``extreme'' precipitation measurements.}
\label{figure5}
\end{center}
\end{figure}

\begin{figure}[!t]
\begin{center}
\includegraphics[trim={0 0 0 0mm}, clip, width = \textwidth]{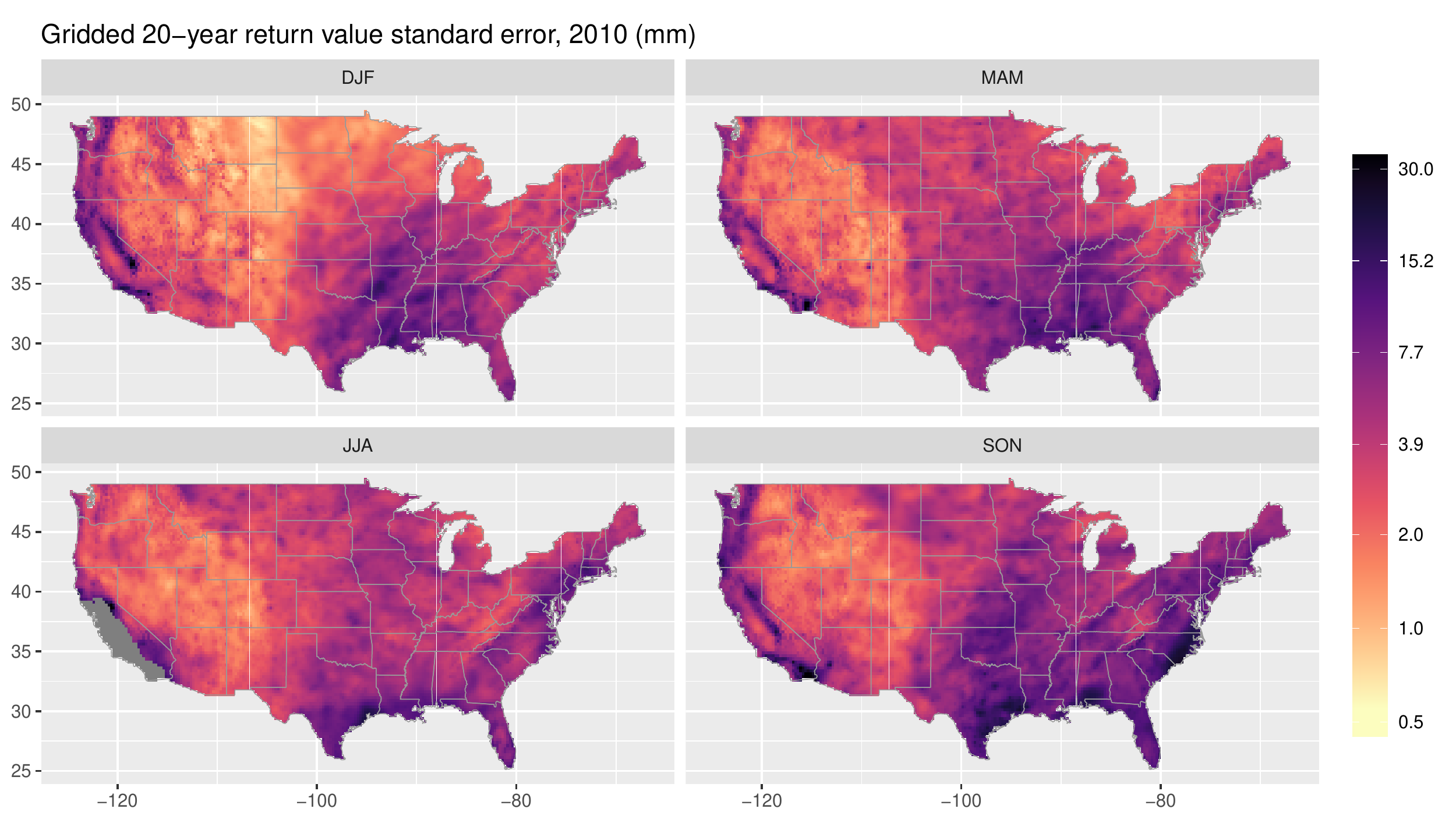}
\caption{Spatially-complete bootstrap standard errors (mm) for the 20-year return value across each season. The standard error provides an estimate of the uncertainty in the return values shown in Figure~\ref{figure5}. Part of California is masked out in JJA due to seasonal dryness (see the caption of Figure~\ref{figure5}).}
\label{figure6}
\end{center}
\end{figure}

Returning to the main goal of the analysis, we now present our new probabilistic gridded product. The product consists of spatially complete maps (on the $0.25^\circ$ grid) of the four climatological coefficients in~(\ref{coef_model}): a best estimate map calculated from the full data, as well as 250 smoothed maps calculated from the bootstrap data sets. With these gridded data in hand, one can create summary statistics for the entire spatial domain as well as bootstrap sampling distributions for confidence intervals or standard errors. Furthermore, these calculations are possible for any individual year in 1950-2017 (note, however, that we have not explicitly accounted for year-to-year variability due to, e.g., ENSO). The most common summaries would likely be return values~(\ref{returnVal}) or return periods~(\ref{returnPer}), but one could just as easily calculate risk ratios for comparing one time point with another (see, e.g., \citealp{Risser2017}) or even trends over time.


As an illustration, Figure~\ref{figure5} shows spatially complete maps of the 20-year return value for each season, and the bootstrap standard errors are shown in Figure~\ref{figure6}.  These maps are unique in that they are high resolution, and they produce return-value estimates consistent with what would be measured at a point (rather than over an area).  As a result, the return value estimates from our method are substantially higher than those estimated from gridded datasets.  For example, Figure A4b in \citet{Volosciuk2015}, shows 20-year DJF return values estimated from the gridded Climate Prediction Center daily precipitation product (the equivalent of the DJF panel in our Figure~\ref{figure5}), and the highest daily return values shown are only about 125~mm, in comparison to values approaching 200~mm estimated using our method. 
We explore this comparison more deeply in Section~\ref{section5}.

The influence of topography, which is explicitly included in the spatial model (see Section~\ref{ssec:spatialanduq}) is clearly evident throughout Figure~\ref{figure5}, as the Sierra Nevada topographic variability associated with the Basin and Range province is clearly evident: high topography areas generally have larger return values.  Notably, however, the influence of topography in the western U.S.~is substantially damped during the summer convective season, with the Sierra Nevada and Cascade ranges hardly evident.  This variation in orographic influence, which emerges naturally from the data itself, is consistent with what one might physically expect: orographic influence in the western U.S. is most prevalent during the winter season, 
when storm systems have a strong zonal flow that is roughly perpendicular to the dominant orography, and it is weaker during the season when precipitation is primarily convective.  Note that Figure~\ref{figure3} shows that the spatial dependence of DJF extreme precipitation in the western U.S. is more meridional than zonal and appears to be aligned with the dominant orientation of topography in the vicinity.  This would indicate that though winter storms impinging on the western U.S.~have a predominantly zonal flow, extremes tend to spatially co-occur along meridionally-oriented (and somewhat perpendicular) mountain ranges that intercept the flow.
Intriguingly, however, orographic modulation of extremes is also absent in the southwestern U.S., which is the center of action during the monsoon season when large moisture fluxes from the south bring summer precipitation that one might expect to also cause orographically modulated precipitation.
It is plausible that this is because the direction of the incoming moisture flux is roughly parallel, rather than perpendicular, to the dominant direction of ranges in the Basin and Range province, which may weaken the orographic effect.  

Our probabilistic gridded product is freely available via a to-be-determined public repository. The data are packaged together into network common data form (netCDF) files. Three separate files are available, with data provided separately for each season: (1) best estimates of the GEV coefficients (e.g., Figure~\ref{figure2}a) as well as bootstrap standard errors, (2) 10-, 20-, 50-, and 100-year return values (with bootstrap standard errors) for 1955, 1965, 1975, 1985, 1995, 2005, and 2015, and (3) smoothed bootstrap estimates (250 total) for each GEV coefficient (these are used to calculate the bootstrap standard errors in the first two files). We also provide code that can be used to calculate return values, return periods, or any function thereof for any year in 1950-2017, as well as the corresponding bootstrap sampling distribution (which can be used to quantify uncertainty via standard errors or confidence intervals).

\subsection{Comparison to the Livneh data product} \label{section5}

As mentioned in Section~\ref{section1}, gridded data products are often used in place of station data to calculate the extreme statistics of precipitation. In light of our novel approach for calculating extreme statistics over CONUS, we now compare our results with a more traditional analysis using the Livneh gridded data product (\citealp{Livneh2014}). Using this data product, we conduct the extreme value analysis described in Section~\ref{section31} independently for each grid cell over CONUS, using the block bootstrap (but without spatial smoothing) to calculate uncertainties.

\begin{figure}[!t]
\begin{center}
\includegraphics[trim={0 0 0 0mm}, clip, width = \textwidth]{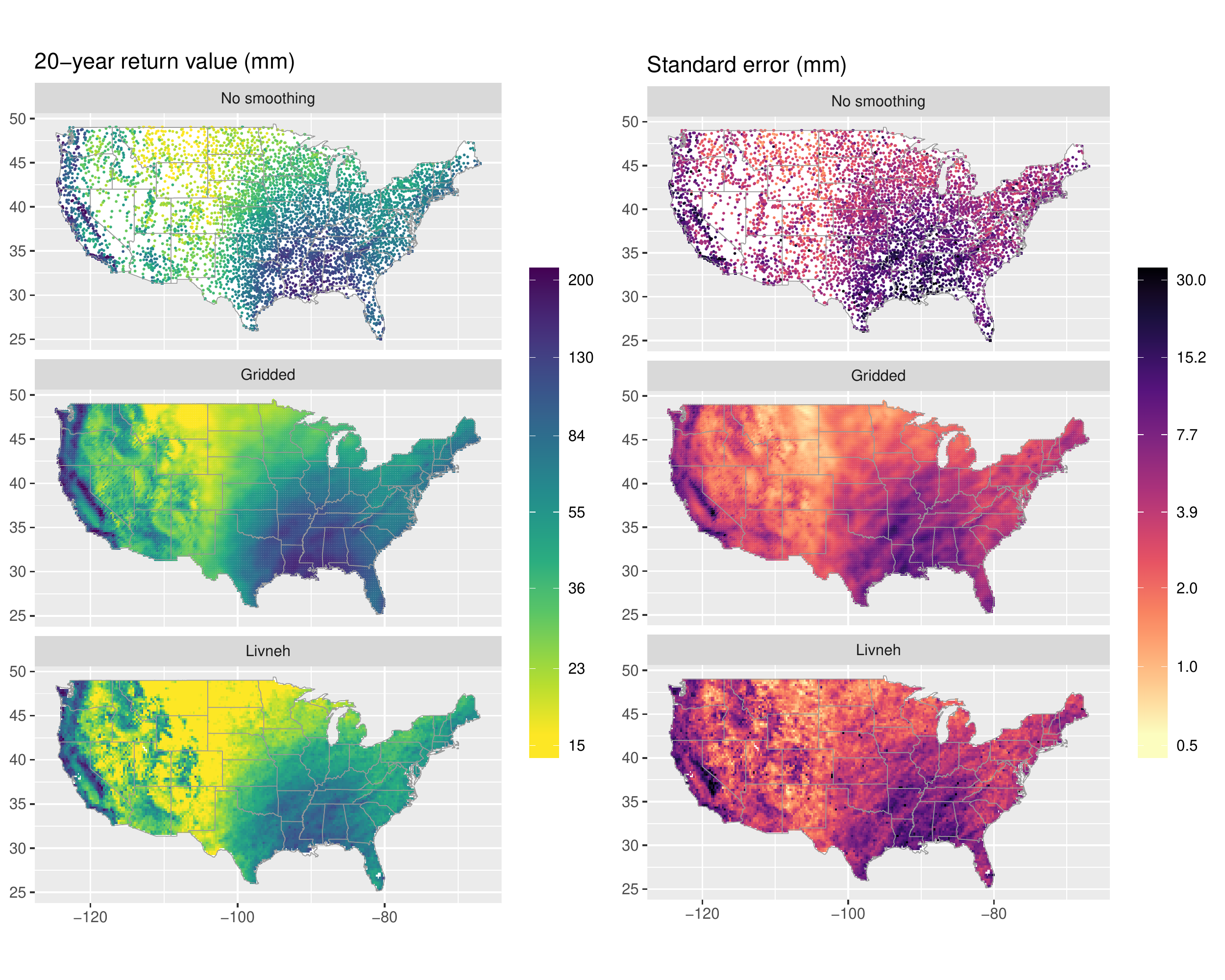}
\caption{DJF 20-year return values for 2010 (left) and standard errors (right) for an analysis with no smoothing (top), our gridded product (middle), and for an analysis using the Livneh data product (bottom). }
\label{figure78}
\end{center}
\end{figure}

\begin{figure}[!t]
\begin{center}
\includegraphics[trim={0 0 0 0mm}, clip, width = \textwidth]{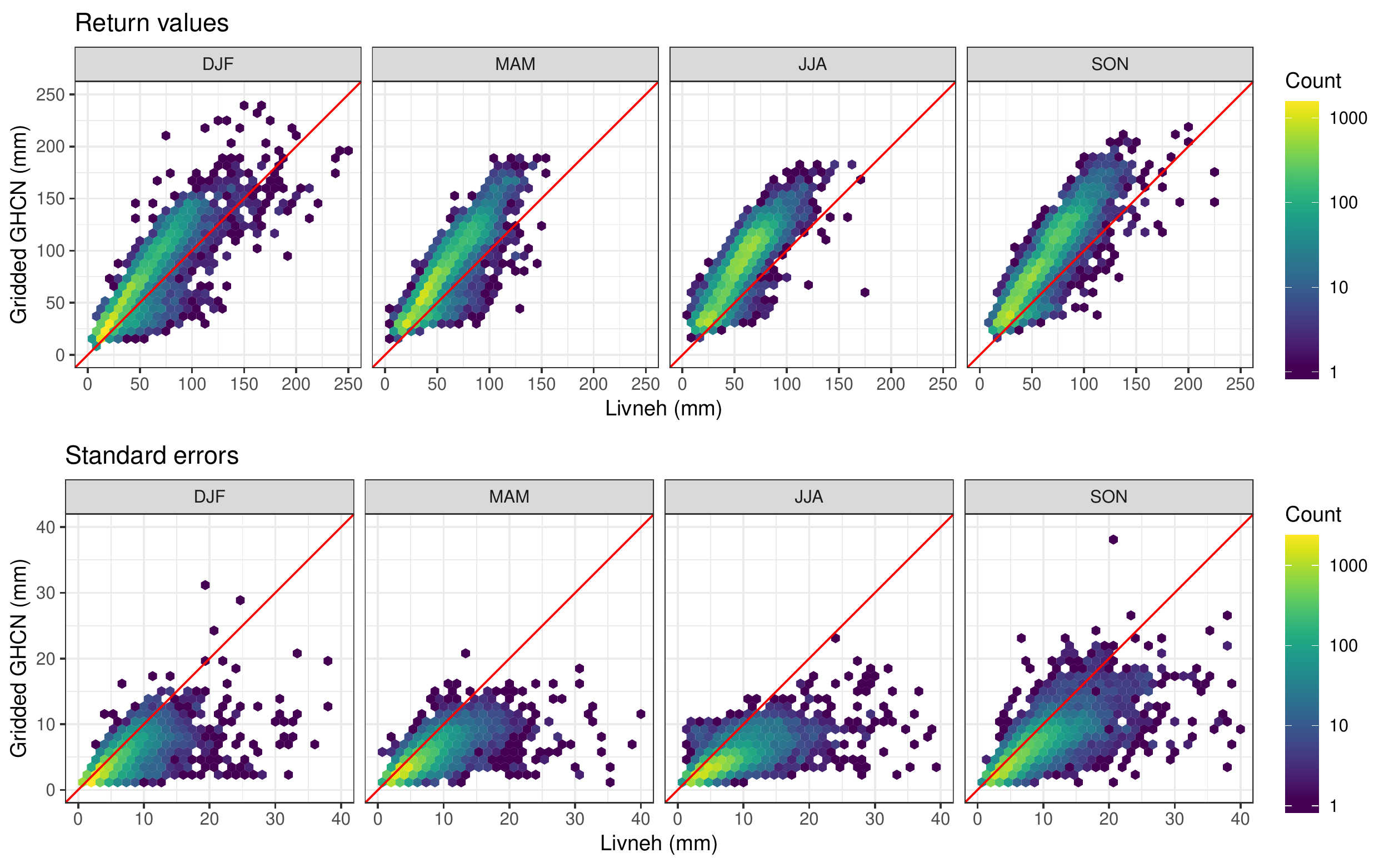}
\caption{Seasonal density scatter plots of return values and their standard errors, with quantities from the Livneh data product on the $x$-axis and corresponding quantities from our GHCN station data analysis on the $y$-axis. The red line indicates the $45^\circ$ line.}
\label{figure9}
\end{center}
\end{figure}

A comparison of the DJF 20-year return values and standard errors for 2010 is shown in Figure~\ref{figure78}: this plot shows estimates without spatial smoothing (also shown in Figure~\ref{figure1}) and our gridded product, as well as estimates based the Livneh data product. Livneh return values and standard errors for other seasons are show in Figures \ref{figure7} and \ref{figure8}, which can be directly compared with Figures \ref{figure5} and \ref{figure6}. A visual comparison of the plots in Figure~\ref{figure78} reveals two immediate points: 
(1) the return values calculated from the Livneh data product are systematically smaller than the return values generated using the station data, both smoothed and raw, and (2) the standard errors from both approaches appear to be approximately the same. In other words, our gridded product produces return values that are consistent with station estimates but higher than estimates from gridded data, while reducing uncertainty relative to independent station estimates. Figure~\ref{figure9} shows a direct comparison of these two quantities by plotting our gridded estimates versus the corresponding Livneh estimates: in the top row, note that the station data return values are uniformly larger than the Livneh return values, and that the bias is worse for the largest return values.  In the bottom row of Figure~\ref{figure9}, however, we can see that the scatterplots are clustered around the $45^\circ$ line, meaning that the standard errors are comparable. There is, however, a tendency for the Livneh standard errors to be larger than the GHCN standard errors, particularly for locations with greater uncertainty. And, since the Livneh return values are smaller than the GHCN values, this tendency for large Livneh standard errors is increased when they are expressed as percentages. 

\section{Conclusions} \label{section6}

In this paper, we have developed novel statistical methodology for conducting a spatial analysis of extreme precipitation for a large network of weather stations over a heterogeneous domain. Using Gaussian processes, our approach uses data-driven smoothing to borrow strength over space, which yields increased confidence in our subsequent estimates of the climatology of extreme precipitation. The result of our analysis is a new ``probabilistic'' gridded product specifically designed for characterizing extreme precipitation, which will be made freely available in a public data repository.

Compared to traditional gridded products, our results yield important differences in estimates of return values. Furthermore, our methodology is able to produce spatially-complete, high resolution maps of return values based on irregularly observed station data, and we are therefore able to characterize extreme statistics of precipitation at small spatial scales -- especially relative to the large, spatially-averaged summaries provided in, for example, \cite{Kunkel2013} or \cite{Easterling2017}. As a result, our new gridded product provides important insight into the underlying physics of precipitation over CONUS (both in the spatial length-scale differences and further exploration of the extreme statistics themselves), and the return value maps are of value with respect to impacts of extreme precipitation, resource management, and infrastructure design. 

An important extension of the results presented in this paper involves characterizing high-resolution observed trends in extreme precipitation. In future work we plan to explore the presence and statistical significance of any trends over time in the statistics of extreme precipitation. Such an exploration naturally requires incorporation of more sophisticated trend models where the effects of interannual or decadal variability (e.g., the El Ni\~no/Southern Oscillation or the Pacific Decadal Oscillation) are explicitly accounted for. However, introducing additional time-varying covariate information into a statistical model for the extreme statistics of precipitation necessitates robust methodology for selecting the variables that are relevant for explaining year-to-year variability.

In conclusion, we note that the correct interpretation and usage of gridded products closely depends upon their construction. In the methodology presented here, the long period return values within a grid box are interpreted as representative of any point within that grid box. A simple interpretation is that for a given return period, any point within the grid box has the same return value. 
On the other hand, modeled daily precipitation is the total precipitation within a grid box integrated at all points within the box as constrained by the conservation properties of the model.  Hence, when compared to return values from a climate model, observed gridded return values obtained by our method  should be considered an upper bound. In other words, if during an extreme storm, simulated precipitation is such that every point within a grid box is simultaneously precipitating at a very high rate then we would expect the modeled return values to agree with our observational estimates. This situation could indeed occur if the model's grid cells are small enough and if storm properties are relatively spatially uniform on this scale. Hence, we might expect that our estimates of mid-latitude winter storm extreme precipitation would agree with the very high resolutions of cloud system resolving models, but not for convective summer storms, which would always be simulated lower than our estimates because of these scale considerations.

Alternatively, one might expect that return values based on gridded daily observed precipitation products, as in \cite{Wehner2013}, might be more directly comparable to modeled precipitation if the gridding process is conservative. However, because the station density is low compared to model grid boxes, the probability of missing an extreme rainfall event with the grid box is high. Hence, return values calculated from a daily gridded precipitation product provide a lower bound on what climate models should be expected to produce. As a model evaluation strategy, we can utilize these two differing ways of estimating observed long period precipitation return values as model performance metrics by considering them as upper and lower bounds on expected model targets. Typically, models at CMIP5-class horizontal resolutions produce seasonal return values that are lower than observational estimate, because simulated gradients of temperature and moisture are weaker than observed. 
As resolution is increased to $\sim$25km, simulated mid-latitude winter extreme precipitation typically compares more favorably with this lower bound, but simulated summer extreme precipitation can be too high, even compared to the upper bounds, due to deficiencies in cumulus convection parameterizations (\citealp{Wehner2014}). It would be reasonable to expect that as higher resolution multi-decadal simulations become available that simulated mid-latitude winter extreme precipitation will be increased but should be less than the lower bound. Improvements in simulated mid-latitude summer extreme precipitation await convection-permitting resolutions.

\begin{acknowledgements}
The authors would like to thank members of the 2018 Statistical and Applied Mathematical Sciences Institute (SAMSI) working group on extreme precipitation (specifically Daniel Cooley, Brook Russell, Richard Smith, Ken Kunkel, and Whitney Huang) as well as Christina Patricola and Benjamin Timmermans for helpful discussions throughout the development of this work. Furthermore, we would like to thank the reviewers of this manuscript for helpful and constructive feedback that improved the quality of the paper.
 
The data supporting this article are based on publicly available measurements from the National Centers for Environmental Information (at \url{ftp://ftp.ncdc.noaa.gov/pub/data/ghcn/daily/}). 

This research was supported by the Director, Office of Science, Office of Biological and Environmental Research of the U.S. Department of Energy under Contract No. DE-AC02-05CH11231 and used resources of the National Energy Research Scientific Computing Center (NERSC), also supported by the Office of Science of the U.S. Department of Energy, under Contract No. DE-AC02-05CH11231.

This document was prepared as an account of work sponsored by the United States Government. While this document is believed to contain correct information, neither the United States Government nor any agency thereof, nor the Regents of the University of California, nor any of their employees, makes any warranty, express or implied, or assumes any legal responsibility for the accuracy, completeness, or usefulness of any information, apparatus, product, or process disclosed, or represents that its use would not infringe privately owned rights. Reference herein to any specific commercial product, process, or service by its trade name, trademark, manufacturer, or otherwise, does not necessarily constitute or imply its endorsement, recommendation, or favoring by the United States Government or any agency thereof, or the Regents of the University of California. The views and opinions of authors expressed herein do not necessarily state or reflect those of the United States Government or any agency thereof or the Regents of the University of California.
\end{acknowledgements}


\begin{appendix} 
\numberwithin{figure}{section}

\section{Exploratory analysis} \label{appendix0}

As described in Section~\ref{section1}, there is clearly spatial dependence in the underlying fields of daily precipitation (due to storm systems; here referred to as ``storm dependence'') that is not accounted for when conducting the GEV analyses on seasonal maxima separately for each station. As such, fields of the pointwise maximum likelihood estimates contain both signal and noise, because the true underlying signal is contaminated by error due to the unaccounted-for storm dependence (note, however, that simple sampling noise or measurement error would occur even without storm dependence). The block bootstrap approach outlined in Section~\ref{section33} can be used to approximate the sampling distribution of parameter estimates at each station and quantify the spatial features of this error. Since we use the same bootstrap samples for all stations, we can also obtain estimates of the pairwise correlations between sampling distributions. Accounting for this correlation allows us to explicitly separate the signal and noise present in the maximum likelihood estimates.

\begin{figure}[!t]
\begin{center}
\includegraphics[trim={0 0 0 0mm}, clip, width = \textwidth]{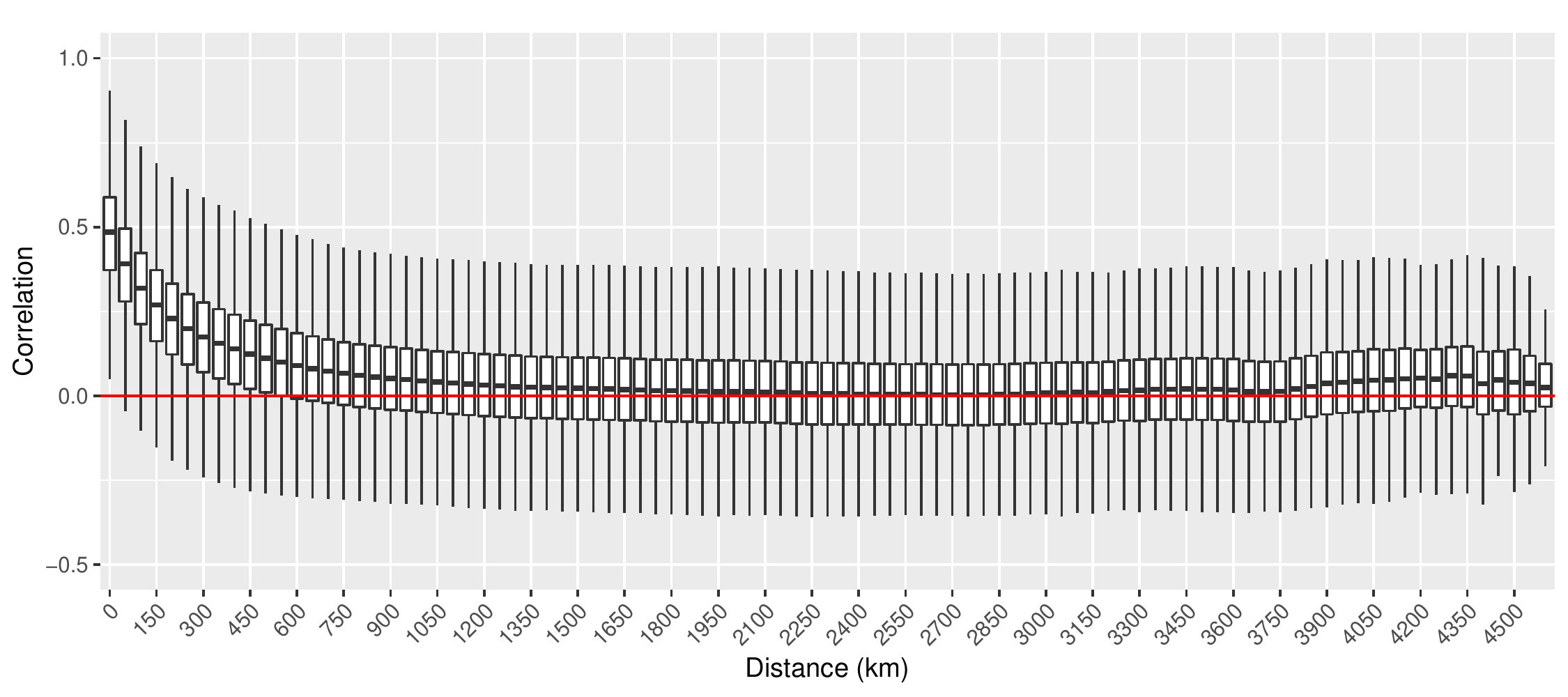}
\caption{Correlogram of bootstrap correlations for the location parameter intercept in DJF.}
\label{correlo_mu0_DJF}
\end{center}
\end{figure}

As a concrete example, consider the following exploratory plots for the location parameter intercept $\mu_0(\bfs)$ for DJF. To explore the co-variance of the pointwise maximum likelihood estimates, we can use the pairwise empirical correlations from the bootstrap samples across all pairs of stations. The easiest way to visualize these correlations simultaneously is via a correlogram, where we bin the pairwise correlations by distance and present box plots for each bin. The correlogram for the location intercept in DJF is shown in Figure~\ref{correlo_mu0_DJF}. Note that there is non-negligible correlation up to approximately 1000km, indicating that the spatial range of correlation in the bootstrap errors is quite large at least for some regions of CONUS. 

\begin{figure}[!t]
\begin{center}
\includegraphics[trim={0 0 0 0mm}, clip, width = \textwidth]{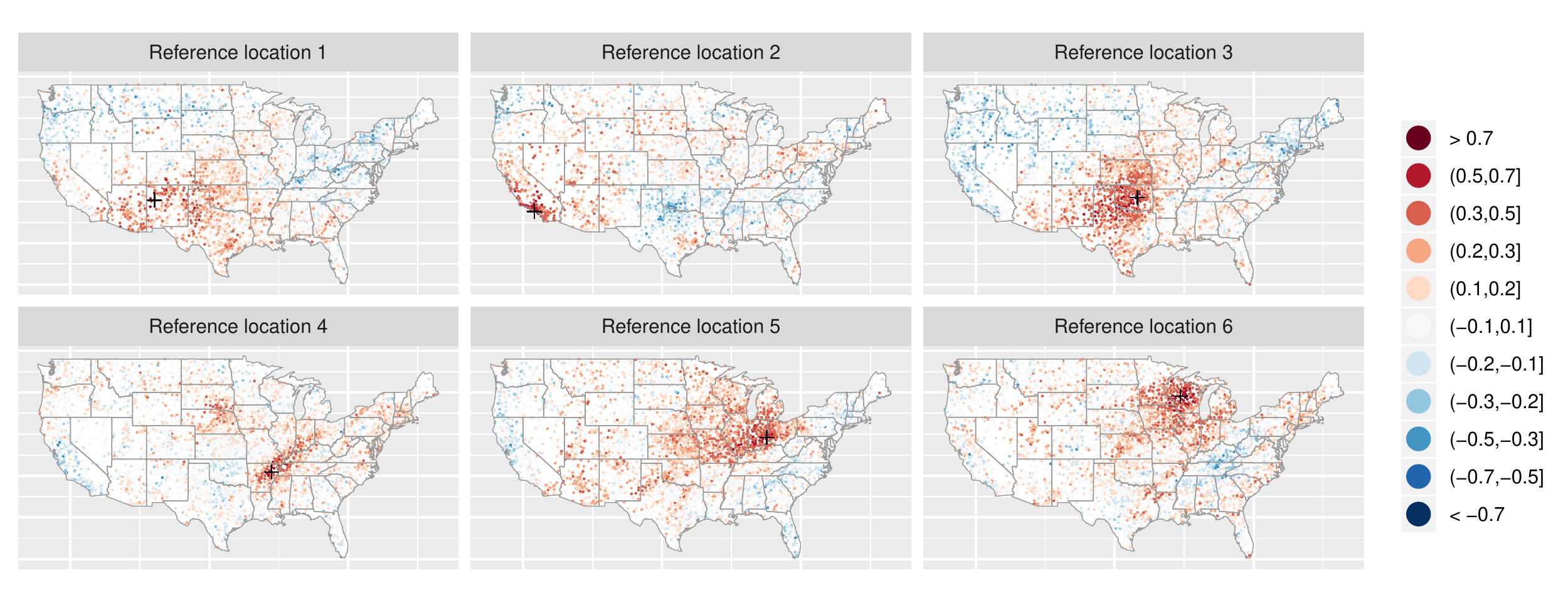}
\caption{Spatial maps of bootstrap correlations for several reference points for the location parameter intercept in DJF.}
\label{bootcorr_mu0_DJF}
\end{center}
\end{figure}

To visualize these correlations more directly for individual stations, consider the plots in Figures \ref{bootcorr_mu0_DJF}, which show spatial maps of the empirical correlations between six selected reference stations and all other stations in CONUS. The plot in the top left of Figure~\ref{bootcorr_mu0_DJF} (Reference location 1) is typical of the correlation map for many stations across CONUS: there appears to be moderate correlations (i.e., approximately 0.5) at very short distances and nonzero correlations extending over a large swatch of the country. The reference station in California (Reference location 2) shows strong local correlation, but this correlation dies off rather quickly. Reference locations 3-6 show very strong local correlations that extend over large, irregular areas. Given that we are considering DJF and Reference locations 3-6 are in the center of the United States, these large clusters of strong correlation are likely the direct result of the winter storm systems that are common in this area. The bottom left plot (Reference location 4) is particularly interesting, as this shows a strong yet narrow band of large correlation, oriented in a southwest-to-northeast direction. Again, while some of these correlations may be spurious, there are definitely non-negligible correlations for many reference locations that extend over large spatial domains.

These explorations confirm the presence of spatially-correlated error in the coefficient estimates at non-negligible scales, and that this noise is a result of the unaccounted-for storm dependence. It is therefore important to account for this error in the second step of the analysis where we estimate the first- and second-order properties of the true spatial signal (Section~\ref{section32}). Unfortunately, it is not entirely clear how to do this, since the spatial coherence of both signal and error appear to be on similar spatial scales. Furthermore, it is obvious that the correlation patterns are highly irregular and long-range (which could be evidence of teleconnections), so we would prefer to use an empirically-based approach rather than complex statistical modeling. The simplest approach would be to use the empirical (sample) covariance matrix of the bootstrap samples as a fixed plug-in estimate for the true error covariance ${\bf D}_\bftheta$, but unfortunately the sample covariance is known to be a poor estimate of the true covariance. Furthermore, since $B=250$ (the number of bootstrap samples)  is less than $n = 5202$ (the number of locations; in practice $B$ could be made arbitrarily large), the sample covariance is singular. Of course, the statistics literature describes a wide variety of strategies for covariance regularization, including modeling the covariance separately in terms of standard deviations and correlations (\citealp{Barnard2000}) and a variety of shrinkage-based approaches (e.g., \citealp{Daniels2001}, \citealp{Schafer2005}, and \citealp{Hannart2014}), among many others. However, it is not immediately clear what type of regularization is appropriate, as some of the long-range correlations shown in Figure~\ref{bootcorr_mu0_DJF} are real. Furthermore, when we implemented a test case where ${\bf D}_\bftheta$ was fixed as the sample bootstrap covariance, we were not convinced that the spatial smoothing/shrinkage was being done correctly (the fact that the signal and error have similar spatial scales may be causing problems). Therefore, as mentioned in Section~\ref{section33}, we used a simple (i.e., diagonal) model for the error covariance estimated directly from the pointwise maximum likelihood estimates, relying on the block bootstrap to account for the storm dependence. Further exploration of this issue is beyond the scope of this paper and will be the subject of future methodological work.

\section{Spatial statistical modeling for the coefficient fields} \label{appendixA}

Recall from Section~\ref{section32} that we account for dependence across the climatological coefficients over the spatial domain using second-order nonstationary spatial Gaussian process models for each of the spatially-varying climatological coefficients in~(\ref{coef_model}). Given the two-stage nature of the analysis, recall from Section~\ref{section33} that we estimate the Gaussian process mean $\boldsymbol{\delta}_\theta$, covariance ${\bf \Sigma}_\theta$, and error covariance ${\bf D}_\bftheta$ using the marginalized model
\begin{equation} \label{app:margmod}
\widehat{\bftheta} \sim N_n(\boldsymbol{\delta}_\theta, {\bf \Sigma}_\theta + {\bf D}_\bftheta),
\end{equation} 
where $\boldsymbol{\delta}_\theta = \big(\delta_\theta(\bfs_1), \dots, \delta_\theta(\bfs_n) \big)$, the $ij$ element of ${\bf \Sigma}_\theta$ is $C_\theta(\bfs_i, \bfs_j)$, and ${\bf D}_\bftheta$ is a diagonal matrix of elements $\tau_\theta^2(\bfs_i)$. In other words, $\tau_\theta^2(\bfs_i)$ quantifies the variability of the error at station $\bfs_i$.

In general, using a Gaussian process to estimate a second-order nonstationary covariance function $C_\theta$ for a large data set is computationally demanding. First of all, for a data set of size $n$, a single evaluation of the multivariate Gaussian likelihood requires $\mathcal{O}(n^2)$ storage and $\mathcal{O}(n^3)$ calculations, which is costly (yet not impossible) for $n = 5202$. More seriously, many nonstationary covariance function models are highly parameterized (we refer the interested reader to \citealp{Risser2016}), and parameter estimation becomes very difficult for large $n$. To navigate both of these difficulties, we elect to use the covariance function outlined in \cite{RisserCalder2017}, which uses local-likelihood estimation and a mixture component technique to estimate the high-dimensional covariance function parameters. The essence of the method is that we first estimate the statistical parameters of~(\ref{app:margmod}) locally and then use a Gaussian smoothing kernel to interpolate the local estimates to the station locations. Formally, the method starts out by defining a coarse grid of $K$ ``mixture component'' locations over the spatial domain, denoted $\{ {\bf b}_k: k = 1, \dots, K\}$. Then, independently for each ${\bf b}_k$, a set of statistical parameters are estimated from a stationary (anisotropic) Gaussian process with a mean that is linear in an elevation-based covariate (as well as an intercept) using data from all of the stations that lie within a particular radius $r$ of ${\bf b}_k$. Let $\phi$ represent an arbitrary parameter in the stationary Gaussian process model, i.e., one of the mean regression coefficients, spatial variance, and spatial covariance parameters (for simplicity, we drop the $\theta$ subscript, but note that these parameters are estimated separately for each climatological coefficient). Then, conditional on mixture component estimates $\{ \widehat{\phi}_k: k = 1, \dots, K\}$, the parameter estimate for an arbitrary location $\bfs$ is
\begin{equation} \label{smoothedPar}
\widehat{\phi}(\bfs) = \sum_{k=1}^K w_k(\bfs) \widehat{\phi}_k,
\end{equation}
where $w_k \propto \exp\{-||\bfs - {\bf b}_k||^2/(2h)\}$ such that $\sum_{k=1}^K w_k(\bfs) = 1$. These estimates are plugged into the nonstationary covariance function outlined in \cite{RisserCalder2017}, which was originally derived in \cite{Paciorek2006} and \cite{Risser2015}, as well as the diagonal error covariance. The estimate for each component of the mean of~(\ref{app:margmod}) is
\begin{equation} \label{eq:spatMean}
\widehat{\delta}(\bfs) = \widehat{\beta}_0(\bfs) + \widehat{\beta}_1(\bfs) x(\bfs),
\end{equation}
where $x(\bfs)$ is an elevation-based covariate at location $\bfs$ and $\widehat{\beta}_0(\bfs)$ and $\widehat{\beta}_1(\bfs)$ are estimated as in~(\ref{smoothedPar}) from the local coefficient estimates $\{ \widehat{\beta}_0^k, \widehat{\beta}_1^k: k = 1, \dots, K\}$.

\begin{figure}[!t]
\begin{center}
\includegraphics[trim={0 45 0 10mm}, clip, width = 0.7\textwidth]{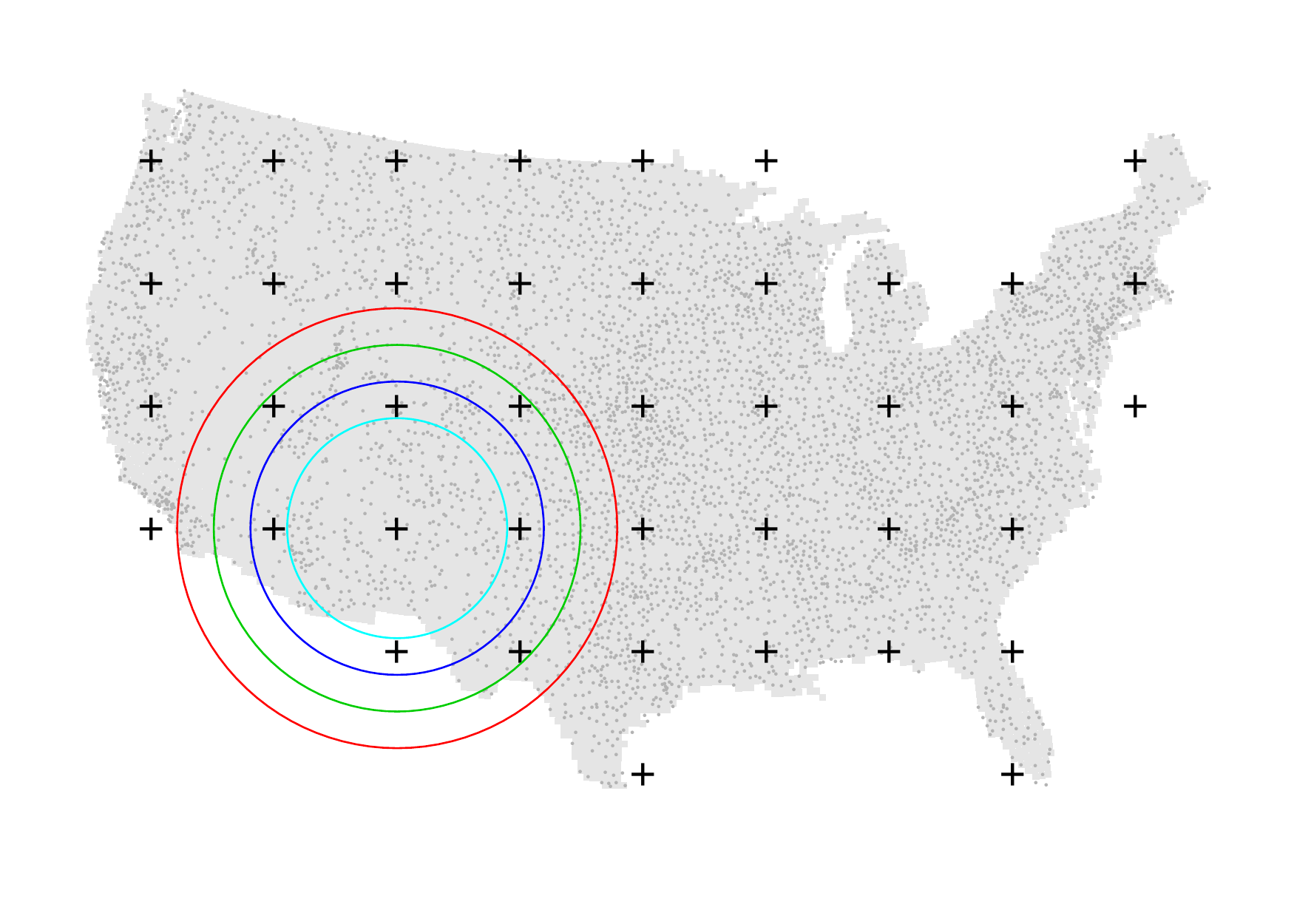}
\caption{The $K=41$ mixture components used for the nonstationary covariance function (black ``+'' symbols), and the four fit radii considered (red = Radius 1, green = Radius 2, blue = Radius 3, and light blue = Radius 4). The GHCN stations used in the analysis are shown with the dark gray dots in the background.}
\label{grid_radii}
\end{center}
\end{figure}

To implement the above statistical model for a given application, however, one must make several choices: 
\begin{enumerate} 
\item A set of mixture component locations $\{ {\bf b}_k: k = 1, \dots, K\}$
\item The local fit radius $r$
\item The bandwidth parameter $h$ in the smoothing kernel
\end{enumerate}
Furthermore, specific to this example, we must also select
\begin{enumerate}  
\item[4.] The local correlation model
\item[5.] An elevation-based covariate for the mean structure
\end{enumerate}
We must make these choices separately for each climatological coefficient $\{\mu_0, \mu_1, \log\sigma, \xi \}$ in each season $\{ \text{DJF, MAM, JJA, SON} \}$. \textit{A priori}, we consider the mixture component locations fixed, using the $K=41$ evenly spaced locations over CONUS shown with the ``+'' symbol in Figure~\ref{grid_radii}, and we also fix the bandwidth parameter to $h = 3$ units, so that the weights at the mixture components themselves, i.e., $w_k({\bf b}_k)$, are approximately 0.95 (on average) and the weights for neighboring mixture component locations are approximately 0.0125 (on average). We would like the data to indicate ``best'' choices for the other selections, as we do not have strong \textit{a priori} expectations about the most appropriate fit radius, local correlation model, and mean covariate. As such, we use 5-fold (out-of-sample) cross-validation and the continuous rank probability score (CRPS) to select the best model, considering all possible combinations of the following:
\begin{itemize}
\item Fit radius: intuitively, a very large fit radius implies that the spatial mean and covariance parameters are very smooth across the spatial domain, while a very small fit radius implies heterogeneity in the parameters. As such, we consider four different fit radii $r_1 = 9$, $r_2 = 7.5$, $r_3 =6$, and $r_4 =4.5$, as these cover a reasonable range of values for defining the local models. Intuitively, these fit radii indicate varying degrees of second-order nonstationarity, with $r_4$ yielding the strongest nonstationarity and $r_1$ yielding the weakest nonstationarity. To compare the fit of a stationary Gaussian process, we also fit a globally stationary model, where all mean and covariance parameters are constant over CONUS (we denote this model as ``radius 0'' or $r_0$).

\item Local correlation: a popular choice in spatial statistics for modeling environmental processes is the Mat\'ern correlation function (see, e.g., \citealp{Stein1999}), which includes a smoothness parameter $\kappa$ that specifically controls the roughness of the fitted surface. We consider two values of $\kappa$: 0.5, which corresponds to the exponential correlation and non-differentiable surfaces, and 2.5, which gives surfaces that are twice-differentiable. 

\item Mean covariate: given the well-established relationships between extreme precipitation and orography, we consider using both elevation and the log of elevation in the Gaussian process mean. 

\end{itemize}
Several other items should be noted: first, we are interested in characterizing large-scale features in the location parameter time trend $\mu_1(\bfs)$ and shape parameter $\xi(\bfs)$, as we expect these features to vary less over space than the location intercept and scale parameters. As a result, for $\mu_1(\bfs)$ and $\xi(\bfs)$ we estimate a locally isotropic correlation (i.e., circular spatial correlation contours), while for $\mu_0(\bfs)$ and $\log\sigma(\bfs)$ we estimate a locally anisotropic correlation (i.e., elliptical spatial correlation contours). Second, given that the true time trend $\mu_1(\bfs)$ is likely to have only large scale features, we only consider the $\kappa=2.5$ model for $\mu_1(\bfs)$.

\begin{table}[!t]
\caption{Best spatial models for each season, in terms of the continuous rank probability score for out-of-sample cross validation.}
\begin{center}
\begin{tabular}{|c|c|c|c|c|}
\hline
Season & Coefficient & Radius & Mean covariate & Smoothness \\ 
\hline \hline
DJF & $\mu_0$ & 4 & Elevation & 0.5\\
\hline
DJF & $\mu_1$ & 3 & Elevation & 2.5 (fixed)\\
\hline
DJF & $\log\sigma$ & 4 & Elevation & 0.5\\
\hline
DJF & $\xi$ & 1 & Elevation & 0.5\\
\hline\hline
MAM & $\mu_0$ & 4 & Elevation & 0.5\\
\hline
MAM & $\mu_1$ & 4 & Elevation & 2.5 (fixed)\\
\hline
MAM & $\log\sigma$ & 3 & Elevation & 0.5\\
\hline
MAM & $\xi$ & 3 & Log Elevation & 0.5\\
\hline\hline
JJA & $\mu_0$ & 3 & Elevation & 0.5\\
\hline
JJA & $\mu_1$ & 2 & Elevation & 2.5 (fixed)\\
\hline
JJA & $\log\sigma$ & 4 & Elevation & 0.5\\
\hline
JJA & $\xi$ & 3 & Elevation & 0.5\\
\hline\hline
SON & $\mu_0$ & 4 & Elevation & 0.5\\
\hline
SON & $\mu_1$ & 3 & Log Elevation & 2.5 (fixed)\\
\hline
SON & $\log\sigma$ & 2 & Elevation & 0.5\\
\hline
SON & $\xi$ & 1 & Log Elevation & 0.5\\
\hline
\end{tabular}\end{center}
\label{CRPS_table}
\end{table}%

The best model (in terms of CRPS) for each coefficient and each season is shown in Table \ref{CRPS_table}. While we wish to allow the coefficient in each season to have a unique degree of nonstationarity (i.e., fit radius), we opt to use a fixed mean covariate (elevation) and smoothness ($\kappa=0.5$) across all seasons and coefficients, as these two models are most often chosen as ``best'' (except for the smoothness in $\mu_1(\bfs)$, which is fixed at 2.5). Note that across all seasons and parameters, the nonstationary Gaussian process models (Radii 1-4) are always preferred to the stationary Gaussian process model (Radius 0).

\section{Computational details} \label{appendixB}

The computational resources required for the analysis in this paper are extensive in light of the bootstrap fitting, even given the computational efficiency of the local likelihood method for estimating a nonstationary covariance function. The analysis was feasible in our case thanks to having access to a large supercomputer through the National Energy Research Scientific Computing Center (NERSC). Specifically, we utilized NERSC's newest supercomputer Cori, a Cray XC40 machine comprised of 2388 Intel Xeon ``Haswell'' processor nodes (there are two other node partitions in Cori that we did not use). Each Haswell node consists of 32 cores, with a total of 128 GB DDR4 2133 MHz memory per node.

There are three primary tasks that are time consuming for the analysis. The approximate number of required hours (both wall clock and compute time) and node configuration for each task are given in the following subsections.

\subsection{Station-specific extreme value analysis}

The four-parameter GEV analysis for each station using {\tt climextRemes} (\citealp{R_climextRemes}) -- i.e., estimating $\mu_0$, $\mu_1$, $\sigma$, and $\xi$ -- is actually quite fast; however, recall that there are $5202$ stations and we must conduct the analysis at each station for four seasons and $B=250$ bootstrap samples (plus one more for the full data fits). On the Cori cores, estimating the GEV coefficients for all $5202$ stations takes approximately 15 minutes. Overall, the configuration for this step is as follows:

\begin{itemize}
\item One ``job'' is defined as estimating the GEV coefficients for all $5202$ stations

\item Configuration: the 4 seasons $\times$ (250 bootstrap + 1 full data) $=$ 1004 jobs are spread out over 16 nodes (512 cores)

\item Compute hours: 1004 jobs $\times$ 15 minutes each $\approx$ 250 hours

\item Wall clock (elapsed) hours: approximately 30 minutes total 

\end{itemize}

\subsection{Cross-validation for spatial model selection}

Recall from Appendix \ref{appendixA} that we must fit a set of spatial models to each of (4 seasons) $\times$ (4 GEV coefficients) $\times$ (2 mean covariates) $\times$ (2 smoothness values) $\times$ (5 holdout sets) $=$ 320 models. The benefit of the nonstationary models (radii 1-4) is that we can split each of these fits into $K=41$ local fits, one for each mixture component. The stationary model (radius 0), unfortunately, cannot be divided into any smaller tasks.

Given that the stationary model involves fitting a spatial model to over 5000 stations and the nonstationary model involves fitting local models to sample sizes ranging between 50 and 2000, for computation we split these tasks into two computing jobs:
\vskip2ex

\noindent \textit{Stationary model fitting}
\begin{itemize}
\item One ``job'' is defined as fitting a stationary model to all 5202 stations

\item Configuration: the 320 jobs are spread out over 5 nodes (160 cores)

\item Compute hours: 160 cores $\times$ 10 hours $\approx$ 1600 hours total 
\item Wall clock (elapsed) hours: approximately 10 hours 

\end{itemize}

\noindent \textit{Nonstationary model fitting}
\begin{itemize}
\item One ``job'' is defined as fitting a locally stationary model 

\item Configuration: the 320 combinations $\times$ 4 radii $\times$ 41 local fits $=$ 52480 jobs spread out over 32 nodes (1024 cores)

\item Compute hours: 1024 cores $\times$ 5 hours $\approx$ 5120 hours total 

\item Wall clock (elapsed) hours: approximately 5 hours 

\end{itemize}

\noindent Once the GP parameters have been estimated, it remains to conduct prediction for the held-out data, but this is a relatively low-cost operation computationally, and was conducted on a personal laptop.

\subsection{Spatial model fitting for the full and bootstrap data sets}

The cross-validation is conducted to select a best model for each coefficient/season, shown in Table \ref{CRPS_table}. The next task is to fit the best model to all 5202 stations, for both the full data estimates and the bootstrapped estimates. Given that the best model across all coefficients/seasons is one of the nonstationary models, we can split things up as follows:

\begin{itemize}
\item One ``job'' is defined as fitting a locally stationary model 

\item Configuration: (250 bootstrap + 1 full data) $\times$ 4 coefficients $\times$ 4 seasons $\times$ 41 local fits $=$ 164656 jobs spread out over 64 nodes (2048 cores)

\item Compute hours: 2048 cores $\times$ 6 hours $\approx$ 12288 hours total $\approx$ 1.4 years

\item Wall clock (elapsed) hours: approximately 6 hours 

\end{itemize}

\subsection{Kriging for the full and bootstrap data sets}

After estimating the local GP parameters, we need to obtain best estimates for the coefficients (across all seasons and bootstrap data sets) at both the GHCN station locations (5202 locations) and the $0.25^\circ$ grid over CONUS (13073 locations). This step is non-trivial, as it involves computing a $5202 \times 5202$ covariance matrix and a $13073 \times 5202$ cross-covariance matrix, as well as computing the Cholesky decomposition of a $5202 \times 5202$ matrix and the subsequent large matrix operations (see Equation \ref{predMean} in the main text). Given the large memory requirements of these calculations (recall that for a data set of size $n$, a single evaluation of the Gaussian likelihood requires $\mathcal{O}(n^2)$ storage and $\mathcal{O}(n^3)$ calculations), the following calculations were done using an alternative computing cluster (2x Eight Core Xeon E5-2680 2.7GHz machine with 384GB total memory):

\begin{itemize}
\item One ``job'' is defined as calculating the kriging predictor for 5202 + 13073 locations

\item Configuration: (250 bootstrap + 1 full data) $\times$ 4 coefficients $\times$ 4 seasons $=$ 4004 jobs spread out over 16 cores

\item Compute hours: 4004 jobs $\times$ 6.2 minutes each $\approx$ 410 hours

\item Wall clock (elapsed) hours: approximately 17 hours 

\end{itemize}

\subsection{Total time}

Aggregating all of the above:

\begin{itemize}
\item Compute hours: $250 + 1600 + 6144 + 12288 + 410 = 20692$ hours

\item Wall clock (elapsed) hours: $0.5 + 10 + 6 + 6  + 17 = 39.5$ hours

\end{itemize}

\clearpage
\section{Supplemental Figures} \label{appendixC}

\begin{figure}[!h]
\begin{center}
\includegraphics[trim={0 27 0 16mm}, clip, width = \textwidth]{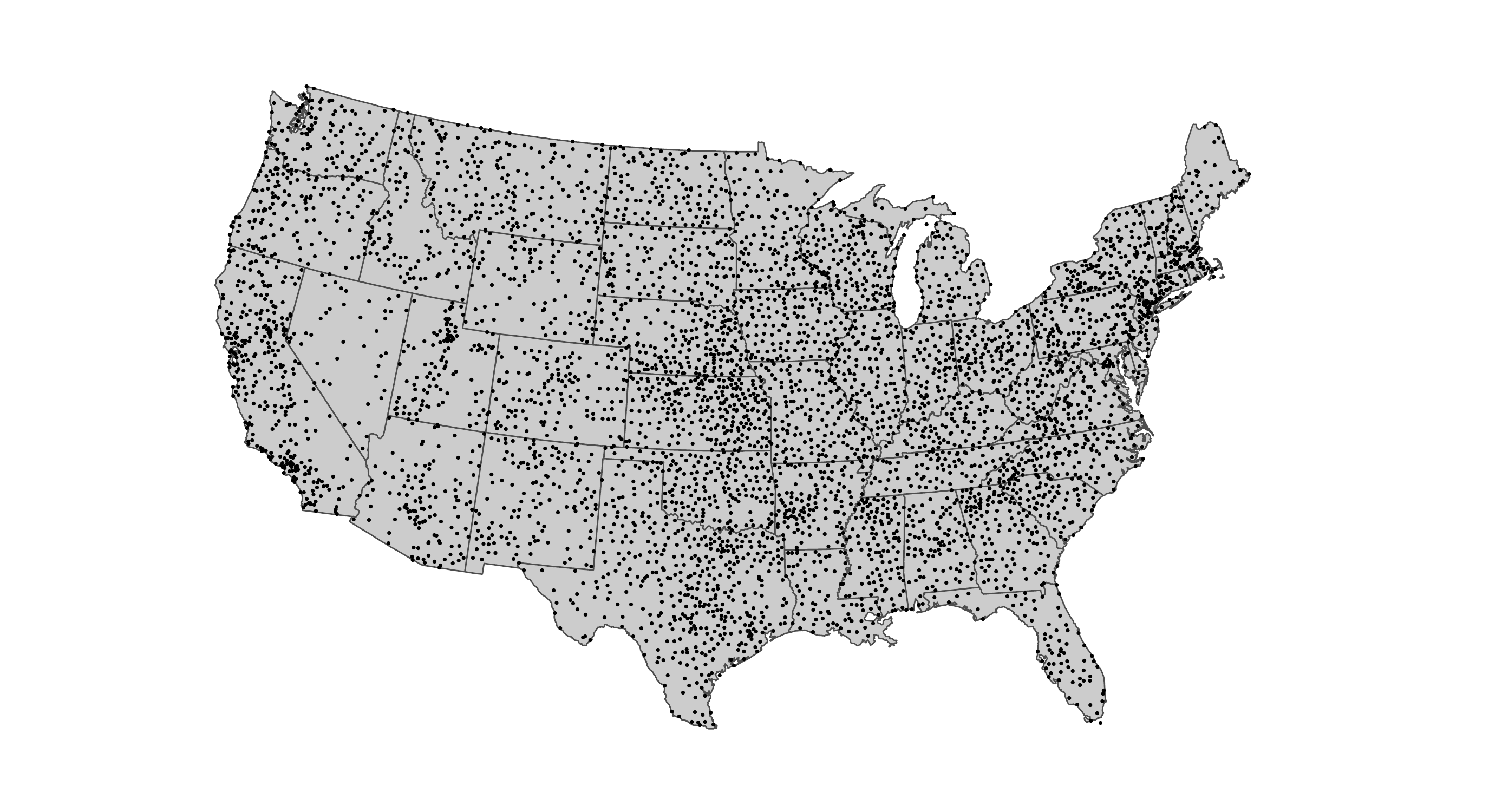}
\caption{The spatial distribution of the $n = 5202$ GHCN stations with a minimum of 66.7\% of nonmissing, quality-controlled daily precipitation measurements during the period spanning December, 1949 to November, 2017.}
\label{GHCNstations}
\end{center}
\end{figure}

\begin{figure}[!t]
\begin{center}
\includegraphics[trim={0 0 0 0mm}, clip, width = \textwidth]{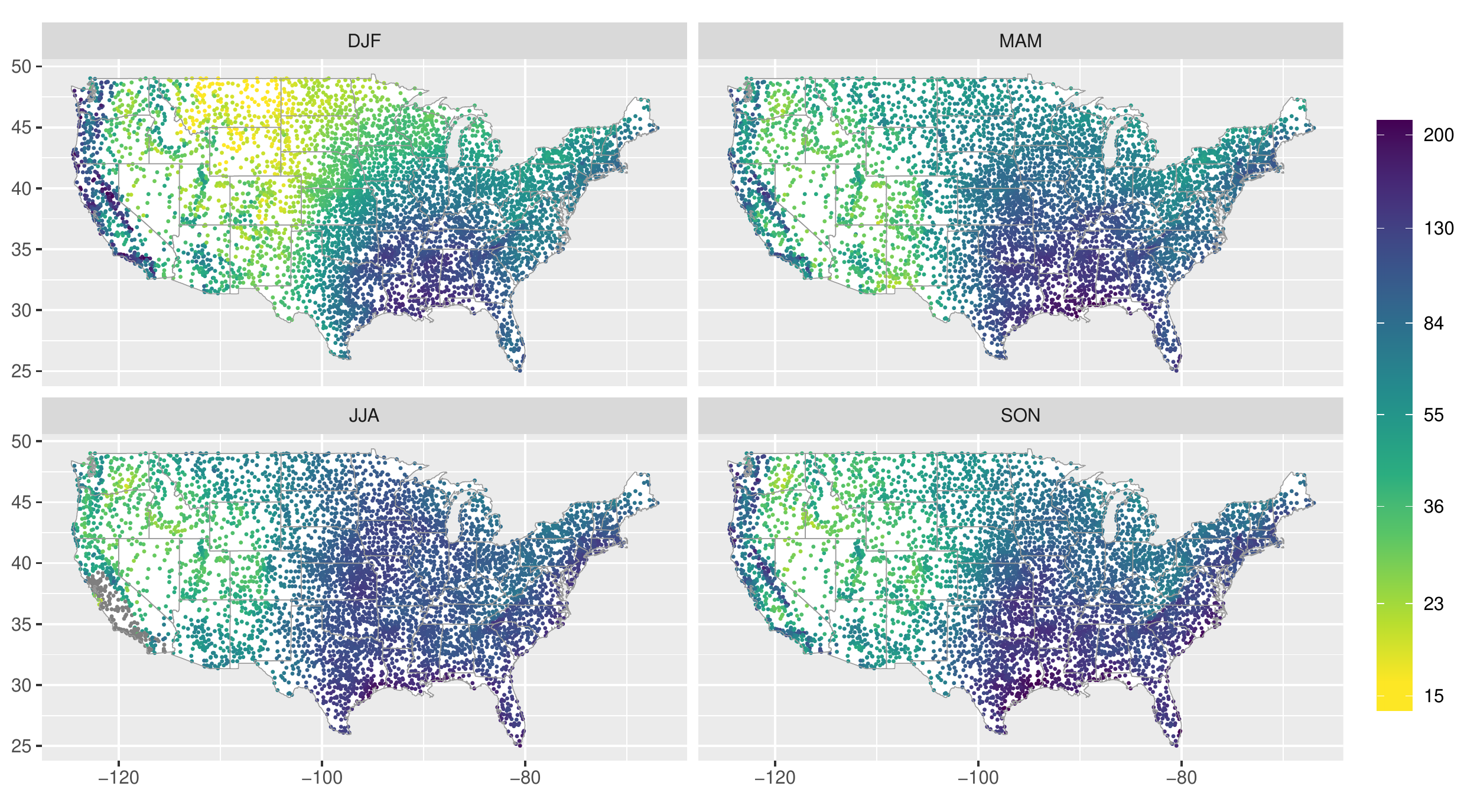}
\caption{Estimated 20-year return values for each season (mm) with spatial smoothing.}
\label{figureC2}
\end{center}
\end{figure}

\begin{figure}[!t]
\begin{center}
\includegraphics[trim={0 0 0 0mm}, clip, width = \textwidth]{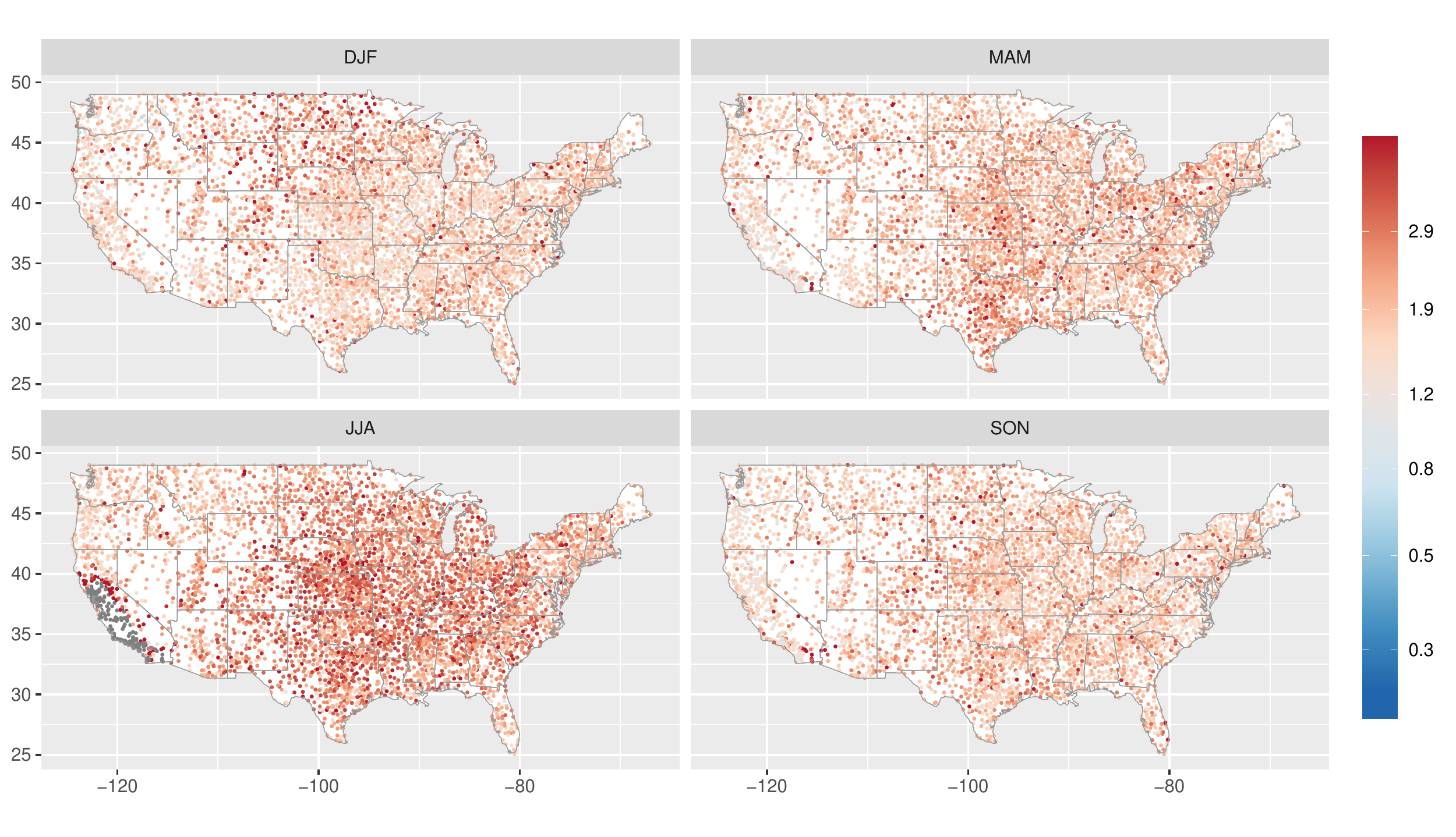}
\caption{Ratio of bootstrap standard errors (no smoothing divided by with smoothing) for each season. Dark red colors indicate stations where the standard errors are much larger without spatial smoothing.}
\label{figureC3}
\end{center}
\end{figure}

\begin{figure}[!t]
\begin{center}
\includegraphics[trim={0 0 0 0mm}, clip, width = \textwidth]{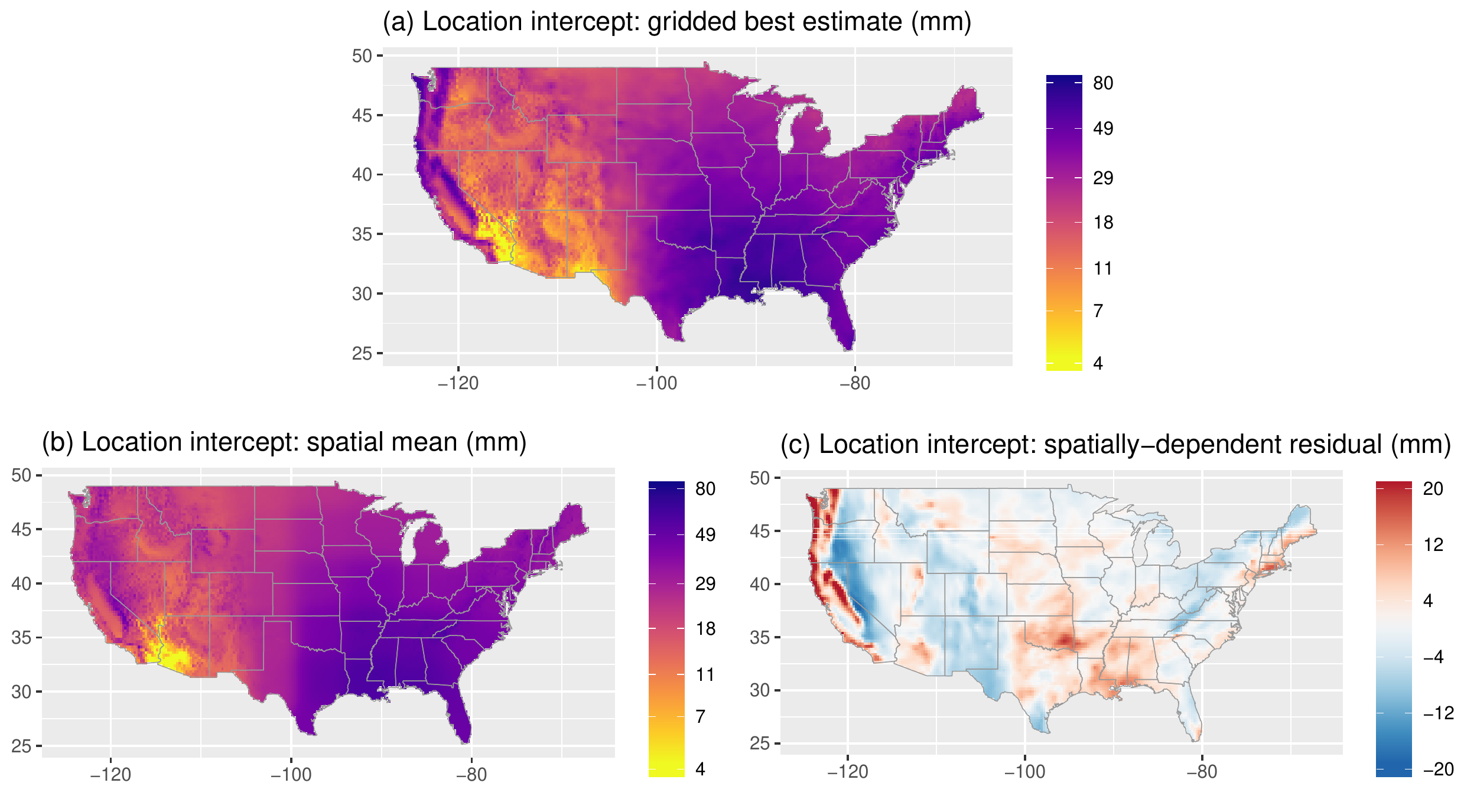}
\caption{Gridded best estimates of the location intercept $\mu_0(\bfs)$ over CONUS (mm) for MAM (panel~a), decomposed into the elevation-based spatial mean (panel~b) and the spatially-dependent residual (panel~c).}
\label{figure2b}
\end{center}
\end{figure}

\begin{figure}[!t]
\begin{center}
\includegraphics[trim={0 0 0 0mm}, clip, width = \textwidth]{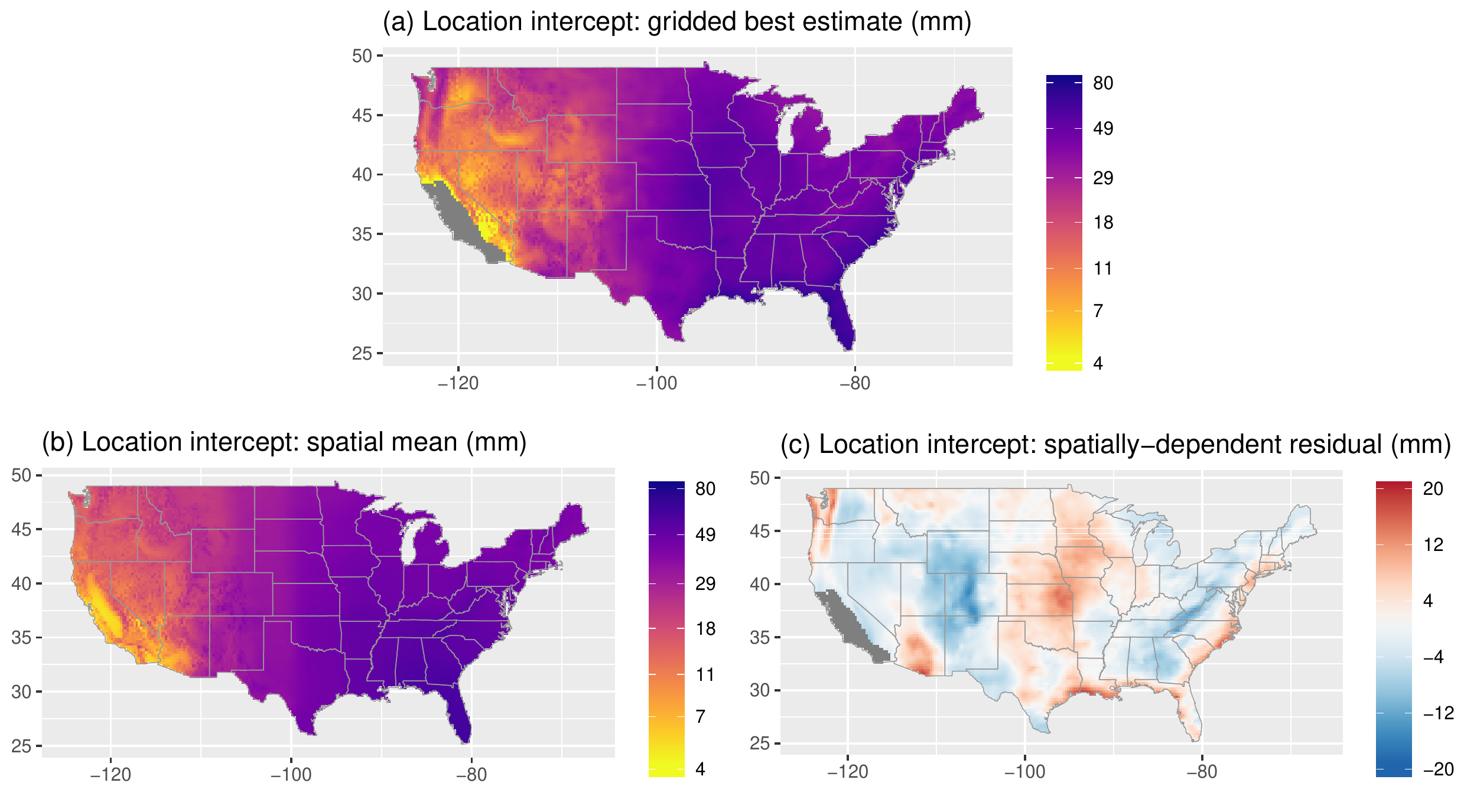}
\caption{Gridded best estimates of the location intercept $\mu_0(\bfs)$ over CONUS (mm) for JJA (panel~a), decomposed into the elevation-based spatial mean (panel~b) and the spatially-dependent residual (panel~c).}
\label{figure2c}
\end{center}
\end{figure}

\begin{figure}[!t]
\begin{center}
\includegraphics[trim={0 0 0 0mm}, clip, width = \textwidth]{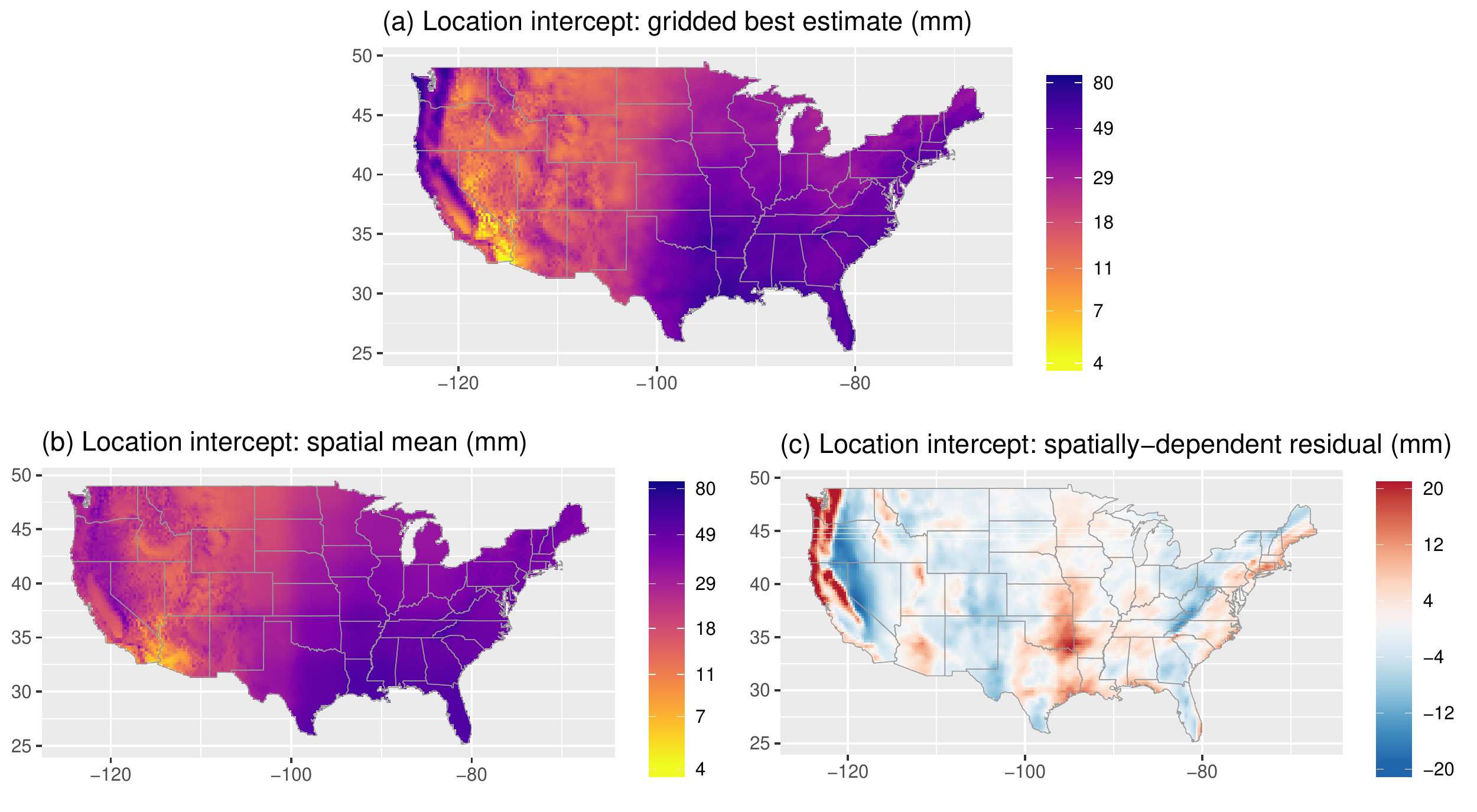}
\caption{Gridded best estimates of the location intercept $\mu_0(\bfs)$ over CONUS (mm) for SON (panel~a), decomposed into the elevation-based spatial mean (panel~b) and the spatially-dependent residual (panel~c).}
\label{figure2d}
\end{center}
\end{figure}

\begin{figure}[!t]
\begin{center}
\includegraphics[trim={0 0 0 0mm}, clip, width = \textwidth]{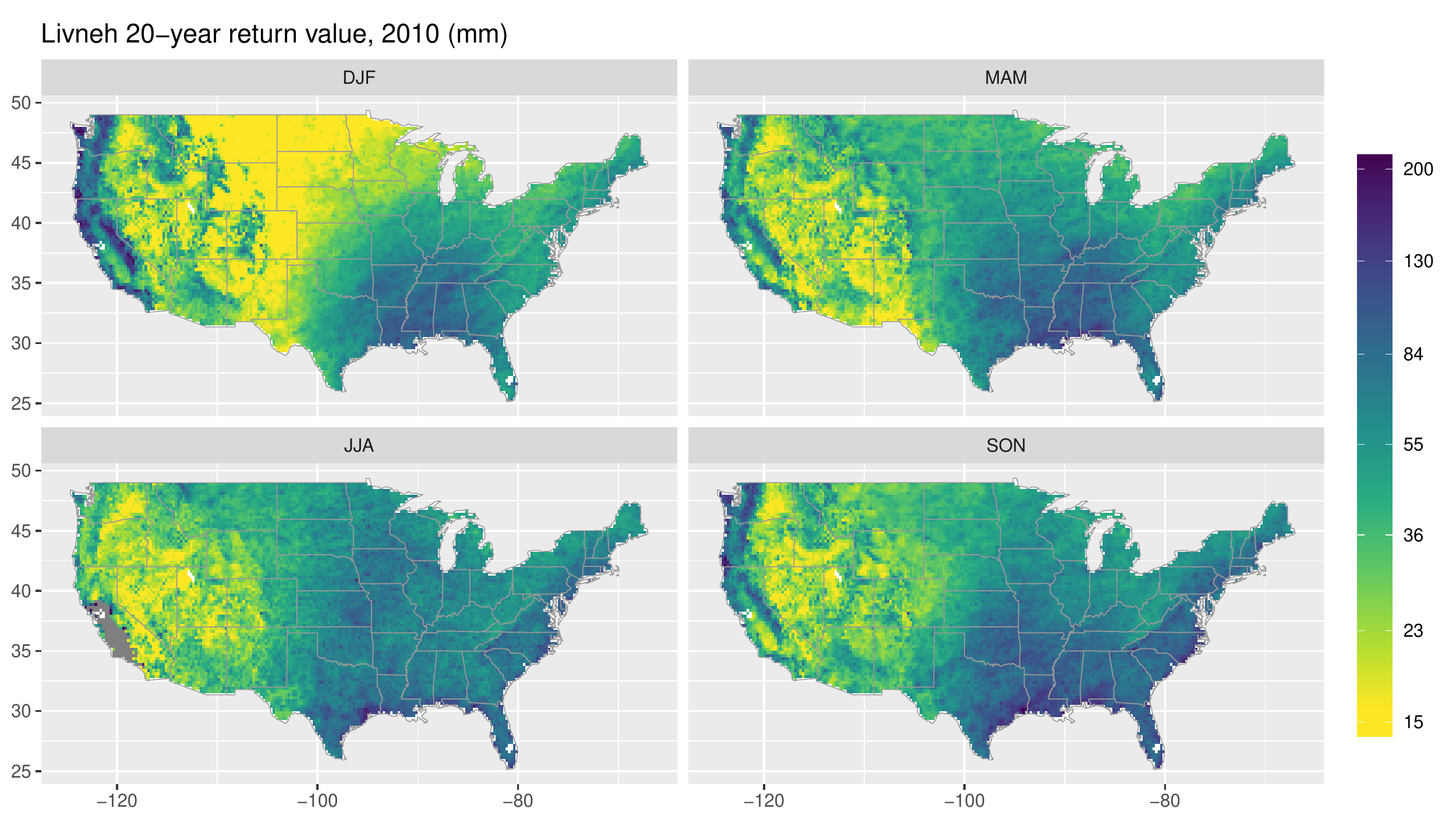}
\caption{20-year return values (mm) calculated from the Livneh data product. This plot is directly comparable with Figure~\ref{figure5}, showing a systematic dampening of the return values.}
\label{figure7}
\end{center}
\end{figure}

\begin{figure}[!t]
\begin{center}
\includegraphics[trim={0 0 0 0mm}, clip, width = \textwidth]{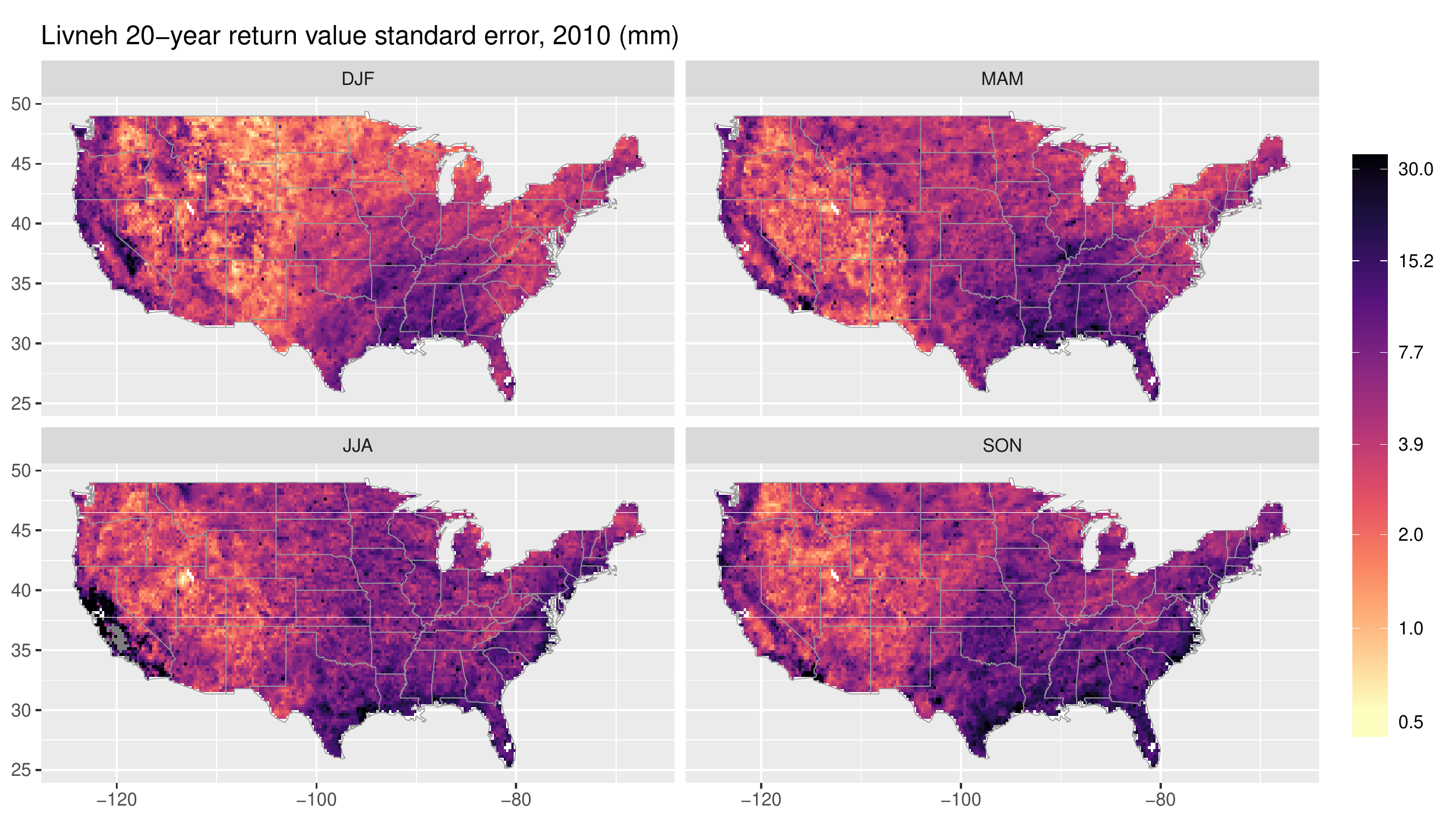}
\caption{Standard errors in the 20-year return values (mm) calculated from the Livneh data product. This plot is directly comparable with Figure~\ref{figure6}, showing general agreement in the standard errors from both approaches.}
\label{figure8}
\end{center}
\end{figure}

\begin{figure}[!t]
\begin{center}
\includegraphics[trim={0 0 0 0mm}, clip, width = \textwidth]{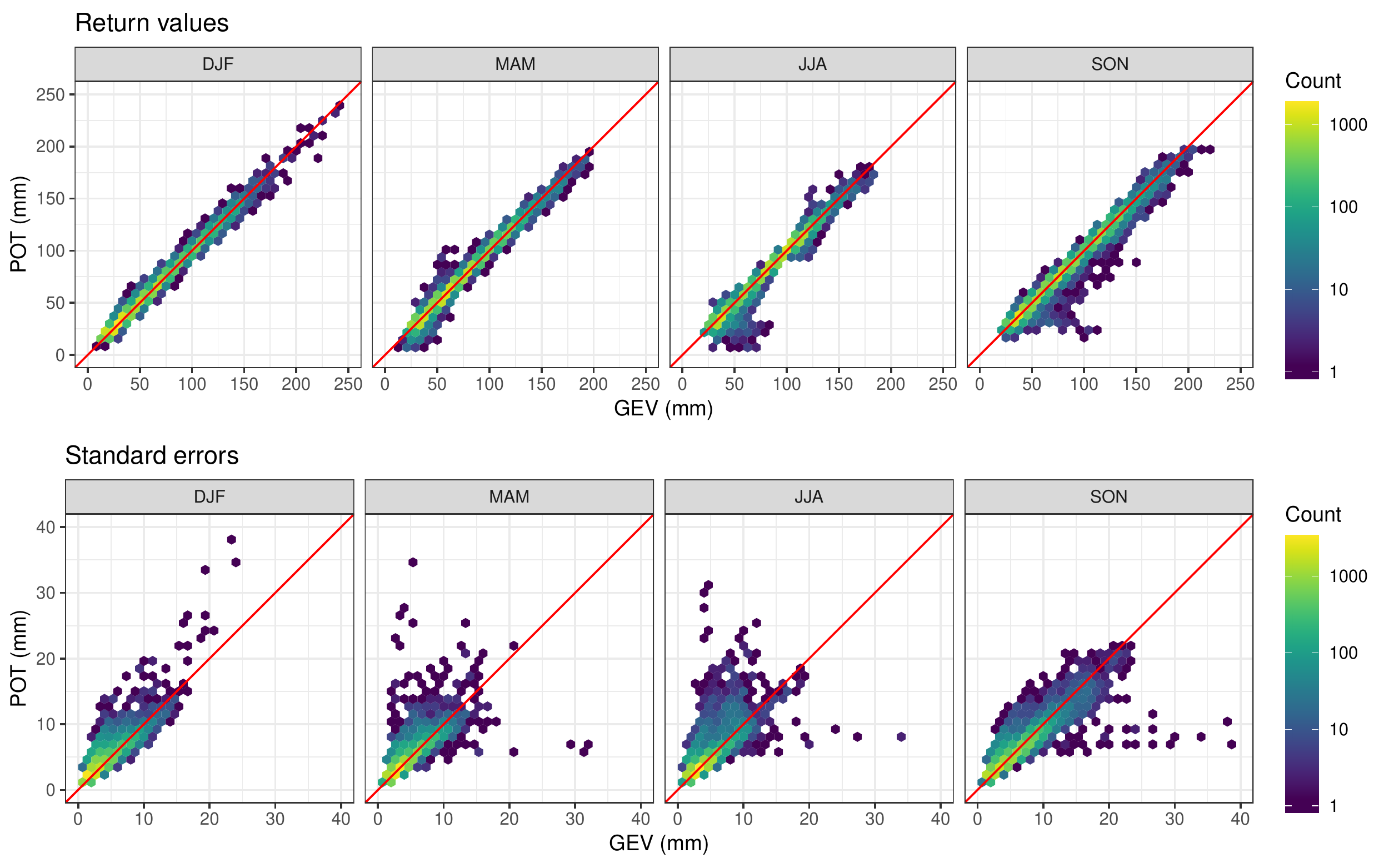}
\caption{A comparison of the 20-year return values (and standard errors) resulting from applying the full two-stage analysis to estimates of the climatological coefficients based on the generalized extreme value (GEV) framework ($x$-axis) and the peaks-over-threshold (POT) framework ($y$-axis). Note that both return value estimates and standard errors are essentially the same for both methods, although the standard errors tend to be slightly larger (on average) for POT vs. GEV.}
\label{figureC4}
\end{center}
\end{figure}

\end{appendix}

\end{document}